\documentclass[12pt]{article}
\pdfoutput=1
\usepackage[a4paper]{geometry}
\usepackage{jheppub,amsmath,amssymb,amsfonts,amsxtra, mathrsfs,makeidx,graphics,graphicx,amsthm,epsfig,bm,longtable,float, color,tikz,mathtools,xfrac,footnote,rotating,lscape,environ,mathtools,empheq}

\usepackage[utf8]{inputenc} 
\usetikzlibrary{positioning}
\usetikzlibrary{chains}
\usetikzlibrary{arrows,fit,decorations.pathreplacing}
\tikzstyle{every picture}+=[remember picture]
\tikzstyle{na} = [baseline]

\usetikzlibrary{arrows, decorations.markings, calc, fadings, decorations.pathreplacing, patterns, decorations.pathmorphing, positioning}

\def\node#1#2{\overset{#1}{\underset{#2}{{\color{gray} \bullet}}}}
\def\Node#1#2{\overset{#1}{\underset{#2}{{ \bullet}}}}

\def\node#1#2{\overset{#1}{\underset{#2}{\circ}}}

\def\blver#1#2{\overset{{\llap{$\scriptstyle#1$}\displaystyle{\color{black} \bullet}{\rlap{$\scriptstyle#2$}}}}{\scriptstyle\vert}}
\def\ver#1#2{\overset{{\llap{$\scriptstyle#1$}\displaystyle\circ{\rlap{$\scriptstyle#2$}}}}{\scriptstyle\vert}}

\tikzstyle{every picture}+=[remember picture]
\tikzstyle{na} = [baseline=-.5ex]

\usepackage{bbm}
\usepackage{subfigure}

\newcommand{\ud}[2]{^{#1}_{\phantom{#1}#2}}

\numberwithin{equation}{section}

\newcommand{\be}{\begin{equation}} \newcommand{\ee}{\end{equation}}
\newcommand{\bea}{\begin{equation} \begin{aligned}} \newcommand{\eea}{\end{aligned} \end{equation}}

\def\tilde{\widetilde}

\def\hat{\widehat}

\def\rt2{\sqrt{2}}

\def\det{\mathop{\rm det}}

\def\tr{\mathop{\rm tr}}

\def\CN{{\cal N}}

\def\1{{\ds 1}}

\def\repa{\raise4pt\hbox{$\square$}\mkern-14mu\raise-4pt\hbox{$\square$}}
\def\repab{\overline{\raise4pt\hbox{$\square$}\mkern-14mu\raise-4pt\hbox{$\square$}\mkern-1mu}}

\def\smileface{\ensuremath{\hbox{\large$\bigcirc$}\mkern-15mu\raise-1pt\hbox{\scriptsize$\smallsmile$}%
\mkern-10mu\raise4pt\hbox{..}\mkern4mu}}
\def\frownface{\ensuremath{\hbox{\large$\bigcirc$}\mkern-15mu\raise-1pt\hbox{\scriptsize$\smallfrown$}%
\mkern-10mu\raise4pt\hbox{..}\mkern4mu}}

\newcommand{\ba}{\begin{array}}
\newcommand{\ea}{\end{array}}
\newcommand{\bi}{\begin{itemize}}
\newcommand{\ei}{\end{itemize}}
\def\vec#1{\bm{#1}}
\def\bea#1\eea{\allowdisplaybreaks \begin{align}#1\end{align}}
 \newcommand{\ben}{\begin{enumerate}}
\newcommand{\een}{\end{enumerate}}
\newcommand{\bean}{\begin{eqnarray*}}
\newcommand{\eean}{\end{eqnarray*}}
\newcommand{\eref}[1]{(\ref{#1})}

\newcommand{\BR}{\mathbb{R}}

\newcommand{\BZ}{\mathbb{Z}}

\newcommand{\comment}[1]{}

\newcommand{\mathd}{\mathrm{d}}
\newcommand{\mathe}{\mathrm{e}}
\newcommand{\mathi}{\mathrm{i}}

\definecolor{light-gray}{gray}{0.7}

\newcommand{\red}{\color{red}}

\newcommand{\Ot}{\mathrm{O3}}
\newcommand{\Ott}{\tilde{\mathrm{O3}}}

\def\aup#1 {\overset{#1}{\uparrow} \, \overset{\tilde{#1}}{\downarrow}}

\title{
Branes,
partition functions and quadratic monopole superpotentials}

\author[a]{Antonio Amariti,}
\author[b]{Luca Cassia,}
\author[c,d]{Ivan Garozzo,}
\author[c,e]{and Noppadol Mekareeya}

\affiliation[a]{INFN, Sezione di Milano, Via Celoria 16, I-20133 Milano, Italy}
\affiliation[b]{
Department of Physics and Astronomy, Uppsala University,
Box 516, SE-75120 Uppsala, Sweden}
\affiliation[c]{INFN, sezione di Milano-Bicocca, Piazza della Scienza 3, I-20126 Milano, Italy}
\affiliation[d]{Dipartimento di Fisica, Universit\`a di Milano-Bicocca,  Piazza della Scienza 3, I-20126 Milano, Italy}
\affiliation[e]{Department of Physics, Faculty of Science, 
Chulalongkorn University, Phayathai Road, 
Pathumwan, Bangkok 10330, Thailand}

\emailAdd{antonio.amariti@mi.infn.it}
\emailAdd{luca.cassia@physics.uu.se}
\emailAdd{ivangarozzo@gmail.com}
\emailAdd{n.mekareeya@gmail.com}

\abstract{
We obtain the brane setup describing  3d $\mathcal{N}=2$ 
dualities for $USp(2N_c)$ and $U(N_c)$ SQCD
with monopole superpotentials. 
This classification follows from a complete analysis of affine and 
twisted affine compactifications from 4d.
The analysis leads to a new duality for the unitary case that has been previously overlooked in the literature.
We check this by matching of the three sphere partition function of the two sides of this new duality and find a perfect
agreement.
Furthermore we use the partition function to predict new 3d 
$\mathcal{N}=2$  dualities for  SQCD
with monopole superpotentials and tensorial matter.
}

\begin{document}
\maketitle

\section{Introduction}

In the recent past there has been remarkable progress in the
understanding of 3d dualities with and without supersymmetry.
One of the main roles in the discovery of new dualities and in the 
appearance of new phenomena such as symmetry enhancements
has been played by monopole operators.
The reasons is that these operators can be used to modify the path integral and to constrain
the global symmetries. These constraints give raise to non-trivial IR relations, 
deforming old dualities and generating new ones.
For example this phenomenon has been largely studied in 3d 
$\mathcal{N}=2$ SQCD 
with linear and quadratic monopole superpotentials.

These constructions have led to a series of new results in the last 
decade.
For example linear monopole superpotential were 
used in \cite{Aharony:2013dha}
to explain how to reduce 4d Seiberg duality to 3d
(see also \cite{Niarchos:2012ah} for an earlier attempt). This construction was then 
generalized to theories with more sophisticated gauge and field content in \cite{Aharony:2013kma,Nii:2014jsa,Amariti:2014iza,Amariti:2015vwa,Hwang:2018uyj}.
Moreover the string theory interpretation of this reduction was 
obtained in \cite{Amariti:2015yea,Amariti:2015mva,Amariti:2016kat,Amariti:2017gsm}, by engineering the linear monopole superpotential in terms of D1 branes, along the lines of the construction of \cite{Hanany:1996ie,deBoer:1997ka,Davies:1999uw,Davies:2000nw}.

A similar construction was provided in \cite{Dimofte:2012pd}
to explain the dimensional reduction of
4d $\mathcal{N}=1$ $SU(2)$ SQCD with eight fundamentals.
The presence of a monopole superpotential was crucial in explaining the enhancement of the  $SU(8)$ global symmetry to $E_7$.
By real mass flow it was then shown that there are more general 
types of monopole superpotentials for $U(1)$ theories.
The generalization of this phenomenon to $USp(2N_c)$ with 
an antisymmetric and eight fundamentals was recently discussed in 
\cite{Amariti:2018wht,Benvenuti:2018bav,Fazzi:2018rkr}.
The $U(N_c)$ generalization of the superpotentials introduced in
\cite{Dimofte:2012pd} for the
$U(1)$ models was obtained in 
\cite{Benvenuti:2016wet,Benini:2017dud}.
This construction has been then used in \cite{Benvenuti:2017lle,Benvenuti:2017kud,Aghaei:2017xqe,Agarwal:2018oxb}
to dimensionally reduce  
the 4d $\mathcal{N}=1$ \emph{``Argyres-Douglas Lagrangians"}  discovered in \cite{
Maruyoshi:2016tqk,Maruyoshi:2016aim,Agarwal:2016pjo,Benvenuti:2017bpg}.
Moreover monopole superpotentials  have allowed 
physical interpretation of many mathematical 
identities among hyperbolic hypergeometric integrals \cite{VdB}.
Such identities represent indeed the matching of the three sphere 
partition function between models with monopole superpotentials
turned on. 
Other interesting results involving monopole superpotentials  
have been discussed in  
\cite{Collinucci:2016hpz,Giacomelli:2017vgk, Amariti:2018dat,Aprile:2018oau}

Furthermore some other dualities,  originally conjectured 
in \cite{Benini:2017dud}, involve deformations with 
quadratic monopole operators.
These dualities have been studied extensively in \cite{Amariti:2018gdc}, 
also for the case of real gauge groups.
In this paper we further investigate such dualities, providing two main results:
\begin{itemize}
\item 
\begin{tabular}{|l|}
\hline
We provide the D-brane engineering of the dualities discussed in 
\cite{Benini:2017dud,Amariti:2018gdc}
\\
 involving quadratic monopole superpotentials.
As a bonus we obtain a new \\ 
duality previously overlooked in the literature. \\
\hline
\end{tabular}
\item
\begin{tabular}{|l|}
\hline
We find new dualities with quadratic monopole superpotentials
for $U(N_c)\,\,$ \\ SQCD with and adjoint  
and $USp(2N_c)$ SQCD with an antisymmetric.
 \\
\hline
\end{tabular}
\end{itemize}

\subsubsection*{\it D-brane engineering}

The first part of the paper focuses on the study of 
D-brane setups that reproduce the 3d dualities 
with linear and quadratic monopole superpotentials
for SQCD with unitary and symplectic gauge groups.

Our construction is based on \cite{Amariti:2015yea}:
we consider a brane setup that engineers a 4d theory, 
with a compact space-like direction. 
Typically there are D4, D6 and NS branes in such setups.
In addition O4 and O6 planes are added, in order to 
extend the analysis to the cases with 
real gauge groups and/or tensorial matter.
We perform T-duality along the compact direction and  study the effective 3d models in the T-dual configuration.
The 3d dualities follow from the transition through infinite coupling 
obtained after an opportune move among the NS branes
\cite{Hanany:1996ie}.
Such move modifies the number of D3 branes that engineer the 
gauge sectors of the effective 3d models. 
A common configuration  
corresponds to having stacks of D3 branes  separated along 
the compact direction. 
This separation is associated to the presence of D1 branes, 
that engineer the presence of interactions involving monopole 
operators.
The simplest cases correspond, at the algebraic level, to affine
Dynkin diagrams, and the affine root is associated to a linear
monopole superpotential, usually referred to as the  
Kaluza-Klein monopole superpotential.
The construction has been shown in \cite{Amariti:2016kat}  to reproduce also the linear 
monopole superpotentials introduced in \cite{Dimofte:2012pd} for $U(1)$ models
and then extended in \cite{Benini:2017dud} to the $U(N_c)$ case.

Here we introduce in this description a new ingredient,
in order to reproduce also the dualities with quadratic 
monopole superpotentials discussed in \cite{Amariti:2018gdc}.
It consists in considering compactifications with a twist by an outer automorphism of the gauge algebra.
Eventually we observe that D-branes provide a classification principle for the 3d $\mathcal{N}=2$ dualities with monopole superpotentials.

The general setup is introduced in section \ref{setup}, where we discuss general aspects of the affine and the 
twisted affine algebras.
In section \ref{sp} we discuss the dualities with real gauge groups.
We observe that by considering the affine and the twisted affine compactifications we can reproduce  
the various dualities obtained in \cite{Aharony:1997gp,Aharony:2013dha,Amariti:2018gdc}
for $USp(2N_c)$ gauge theories involving monopole superpotentials.
In section \ref{uni} we consider the case of $U(N_c)$ gauge groups.
In this case we reproduce all the known dualities studied in 
\cite{Benini:2017dud,Amariti:2018gdc}.
Furthermore we obtain a model that has been previously overlooked in the literature.
This corresponds to SQCD with a linear (quadratic) monopole plus a quadratic (linear) anti-monopole superpotential.
As a check we provide the matching of the partition function along the  two sides of this duality.

\subsubsection*{\it Dualities with tensorial matter}

In the second part of the paper, corresponding to section \ref{plaw}, 
we study new 3d $\mathcal{N}=2$  dualities
for  $U(N_c)$  SQCD with one adjoint  
and $USp(2N_c)$ SQCD with one antisymmetric traceless matter field and quadratic monopole superpotential.
In these cases the tensorial matter fields have a power law superpotential, which truncates the chiral ring. Moreover we show that the quadratic monopole superpotentials  are necessarily dressed by powers of the tensorial matter fields. 
We construct the new dualities by modifying the parent dualities obtained 
in \cite{Kim:2013cma,Amariti:2018gdc} for the unitary case and in \cite{Amariti:2015vwa} for the symplectic one.
The deformation corresponds to a quadratic monopole superpotential 
in the electric and in the magnetic phase. This deformation constrains the real masses and the R-charges.
By  studying the effect of this constraint on the equality relating the
partition functions of the parent theories we arrive at a new IR identity.
This new identity corresponds to the matching of the partition functions between the models with quadratic monopole superpotential, which provides a consistency check of the new duality.

\section{The setup}
\label{setup}

\subsection{Twisted compactification and KK monopoles}
Let us consider the reduction of 4d SYM with gauge group $G$ (whose Lie algebra is $g$) on a circle with radius $r$. 
If the boundary condition of the gauge field $A_\mu$ around the circle (say, in the direction $x^4$) is trivial, namely 
\be
A_\mu(x^0, x^1, x^2, x^4+2 \pi r) =  A_\mu(x^0, x^1, x^2, x^4)~, 
\ee
then the expansion of the gauge field into Fourier modes forms the untwisted affine Lie algebra $g^{(1)}$ \cite{Davies:2000nw}.
More generally, one may consider the boundary condition (see \cite{Kim:2004xx} for an extensive discussion):
\begin{equation}
\label{outer}
A_\mu(x^0, x^1, x^2, x^4+2 \pi r) = {\vec \sigma}( A_\mu(x^0, x^1, x^2, x^4))
\end{equation}
where ${\vec \sigma}$ is an outer automorphism of the Lie algebra $g$. For the Lie algebras $g= A_N, \, D_N, \, E_6$, the element $\vec \sigma$ can be of order $L=2$, and for $g=D_4$, $\vec \sigma$ can be of order $L=3$. The Lie algebra $g$ can be decomposed into the direct sum of the eigenspaces $\mathcal{G}_n$ (with $n=0, \ldots, L-1$) associated with the eigenvalues $\mathe^{2 \pi \mathi n/L}$ of $\vec \sigma$:
\be
\vec \sigma(h) = \mathe^{2 \pi \mathi n/L} h~,\qquad \text{for $h \in \mathcal{G}_n$}~.
\ee
The mode expansion of $A_\mu$ can be written as \cite{Kim:2004xx}
\be
A_\mu^i(x^4) T^i = \sum_{m \in \BZ} \sum_{n=0}^{L-1} A_\mu^{i, (m,n)} \exp\left(-\mathi \frac{x^4}{r} \left(m+\frac{n}{L} \right)  \right) T^i = \sum_{m \in \BZ} \sum_{n=0}^{L-1} A_\mu^{i, (m,n)}  T^i_{m+\frac{n}{L}}
\ee
where $T_i$ (with $i=1, \ldots, \dim\, g$) are the {Lie algebra generators}, and we define
\be
T^i_{m+\frac{n}{L}} := \exp\left(-\mathi \frac{x^4}{r} \left(m+\frac{n}{L} \right)  \right) T^i~.
\ee
  
If $\vec \sigma$ is trivial, then $T^a_{m}$ form a set of the generators of the untwisted affine Lie algebra $g^{(1)}$; we refer to this case as an untwisted compactification.  However, if $\vec \sigma$ is non-trivial and is of order $L=2, \,3$, then $T^i_{m+\frac{n}{L}}$ form a set of the generators of the twisted affine Lie algebra $g^{(L)}$; we refer to this case as a twisted compactification.   In the following discussion, we shall focus only on the case of $L=2$, with the twisted affine Lie algebras
$A^{(2)}_{2N-1}$, $A^{(2)}_{2N}$, and $D^{(2)}_{N+1}$.

In the three dimensional limit where $r \rightarrow 0$, the gauge algebra $g$ reduces to a smaller Lie algebra $\mathcal{G}_0$, since in this case we do not have a Kaluza--Klein mass term.  The rank $r'$ of $\mathcal{G}_0$ can be smaller than that of $G$.  The simple roots $\beta_a$ (with $a=1,\ldots, r'$) of $\mathcal{G}_0$, together with the lowest negative weight $\beta_0$ of $\mathcal{G}_1$, form the Dynkin diagram of the twisted affine Lie algebra $g^{(L)}$.  

An instanton on $\BR^3 \times S^1$ can be regarded as a composite that contains fundamental monopoles as constituents \cite{Garland:1988bv, Nahm:1982jb,  Gross:1980br, Lee:1997vp, Kraan:1998kp, Lee:1998vu, Hanany:2001iy}.  Each of the fundamental monopole consists of four zero modes, namely three associated with its position and one associated with the phase, and is labelled by the co-root $\beta^*_a$ (with $a=0, \ldots, r'$).  Any other monopole configuration is a composite of such fundamental monopoles.   The aforementioned instanton configuration is characterised by a set of non-negative integers $n_a$ (with $a=0, \ldots, r'$) which count the magnetic charge of each
fundamental monopole $\beta^*_a$.   Such an instanton can contribute non-trivially to the effective potential of the theory.  For example, for the $\CN=1^*$ supersymmetric theory on $\BR^3 \times S^1$ with a twisted boundary condition, the instanton contribution to the holomorphic superpotential is given by \cite{Kim:2004xx}:
\begin{equation}
W 
= 
\frac{2}{\beta_0^2} \eta \mathe^{\frac{4 \pi \mathi \tau}{L \beta_0^2}+ \beta_0^*\cdot X} 
+ 
\sum_{a=1}^{r'} \frac{2}{\beta_a^2}  \mathe^{\beta_a^*\cdot X} 
\end{equation}
where $X$ is the adjoint chiral field in the theory and $\tau$ is the holomorphic coupling of the theory.

\subsection{Brane configurations}
From the string theory perspective, the instanton configuration discussed above can be realised from the brane system containing D0 and D4 branes, possibly with the presence of the orientifold fourplane, where fourbranes span $\BR^3 \times S^1$.  Upon using $T$ duality along the $S^1$ direction, we obtain the system consisting of D1 branes stretching between D3 branes, possibly with the presence of orientifold threeplanes. 
The T-dual radius is $R = \frac{\alpha'}{r}$.
 The effect of T-duality on various types of orientifold fourplane is tabulated below.
\be
\begin{tabular} {|c|c|}
\hline
Orientifold & T-duality  \\
\hline
$\widetilde{\mathrm{O4}}^-$ & $\Ot^-$ \& $\Ott^-$  \\
$\mathrm{O4}^+$ & $\Ot^+$ \& $\Ot^+$  \\
$\mathrm{O4}^-$ & $\Ot^-$ \& $\Ot^-$  \\
$\widetilde{\mathrm{O4}}^+$ & $\Ott^+$ \& $\Ot^+$  \\
\hline
\end{tabular}
\ee

The brane configurations that give rise to the untwisted affine Dynkin diagrams as quiver gauge theories on the D1 branes are tabulated below \cite{Hanany:2001iy} (see also \cite{Cremonesi:2014xha}).  The affine node is denoted in black.
\begin{longtable}{|c|c|c|}
\hline
$g$  & Untwisted affine Dynkin diagram of $g^{(1)}$ & Brane set-up\\
\hline
$A_{N}$ & $\begin{array}{l} 
\raisebox{-12pt}{\rotatebox{30}{$-\!\!-\!\!-$}}\node{}{k}\raisebox{0pt}{\rotatebox{-30}{$-\!\!-\!\!-$}} \\[-7pt]
\node{}{k}-\node{}{k}\cdots-\Node{}{k} \quad {\footnotesize \text{($N$ nodes)}}
\end{array}$ &  \begin{tabular}[c]{@{}c@{}} \\ \begin{tikzpicture} [scale=0.9, transform shape]
\draw (0,0.4)--(0,2);
\draw (1,0)--(1,1.6);
\draw (2,0)--(2,1.6);
\draw (3,0)--(3,1.6);
\draw (4,0.4)--(4,2) node[black,midway, yshift=1.2cm] {\footnotesize D3} ;
\draw (0.7,0.9)--(0.7,2.5); 
\draw (1.7,0.9)--(1.7,2.5); 
%
\draw (0,0.9)--(1,0.5) node[black,midway, yshift=0.2cm] {\scriptsize $k$};
\draw (1,0.6)--(2,0.6) node[black,midway, yshift=0.2cm] {\scriptsize $k$};
\draw (2,0.5)--(3,0.5) node[black,midway, yshift=0.2cm] {\scriptsize $k$};
\draw (3,0.6)--(4,0.9) node[black,midway, yshift=0.2cm] {\scriptsize $k$} node[black,midway, xshift=-0.1cm, yshift=-0.3cm] {\scriptsize D1} ;
\draw (0,1.2)--(0.7,1.5) node[black,midway, yshift=0.2cm] {\scriptsize $k$};
\draw (0.7,1.7)--(1.7,1.7) node[black,midway, xshift=0.2cm, yshift=-0.2cm] {\scriptsize $k$}; 
\draw [loosely dotted,rounded corners=0.95cm] (1.7,1.8)--(3.2,1.8)--(4,1.4);
%
\draw [decorate, decoration={brace, mirror}](0,-0.1)--(4,-0.1) node[black,midway,yshift=-0.5cm] {\footnotesize $N~\text{intervals}$};
\end{tikzpicture} \end{tabular} \\
$B_{N}$   & $\Node{}{k}-\node{\ver{}{k}}{2k}-\underbrace{\node{}{2k}-\cdots-\node{}{2k}}_{N-3~\text{nodes}}\Rightarrow\node{}{k}  \ \tikz[na]\node(B1){}; $ &  \begin{tabular}[c]{@{}c@{}} \\  \begin{tikzpicture} [baseline=0, scale=0.9, transform shape]
\draw [ultra thick] (0,0)--(0,2.5) node[black,midway, xshift =-0.3cm, yshift=-1.5cm] {\footnotesize $\Ot^-$}; 
\draw (0.5,0)--(0.5,2.5); \draw (1,0)--(1,2.5); \draw (1.5,0)--(1.5,2.5); \draw (3,0)--(3,2.5); \draw (3.5,0)--(3.5,2.5); \draw (4,0)--(4,2.5); 
\draw [ultra thick] (4.5,0)--(4.5,2.5) node[black,midway, xshift =0.3cm, yshift=-1.5cm] {\footnotesize $\Ott^-$};
%
\draw [thick, color=red, rounded corners=0.75cm](0.5,1)--(-0.4,0.75)--(1,0.75) node[black,midway, xshift=-0.0cm, yshift=-0.2cm] {\scriptsize $k$} ;
%
\draw (0.5,0.9)--(1,0.9) node[black,midway, yshift=0.4cm] {\scriptsize $k$};
\draw (1,1.3)--(1.5,1.3) node[black,midway, yshift=0.4cm] {\scriptsize $2k$} ;
\draw (1,1.5)--(1.5,1.5) node[black,midway, yshift=-0.4cm] {\tiny D1} ;
\draw [loosely dotted] (1.5,1)--(3,1);
\draw (3,1.3)--(3.5,1.3) node[black,midway, yshift=0.4cm] {\scriptsize $2k$};
\draw (3,1.5)--(3.5,1.5) node[black,midway, xshift =0.3cm, yshift=1.4cm] {\footnotesize D3} ;
\draw (3.5,1.7)--(4,1.7) node[black,midway, yshift=0.5cm] {\scriptsize $2k$};
\draw (3.5,1.9)--(4,1.9);
%
\draw [thick, color=red, rounded corners=0.75cm](4,0.75)--(4.9,0.75) --(4,1) node[black,midway, xshift=-0.2cm, yshift=0.3cm] {\scriptsize $k$};
%
\draw [decorate, decoration={brace, mirror}](1,-0.1)--(4,-0.1) node[black,midway,yshift=-0.5cm] {\footnotesize $N-2~\text{intervals}$};
\end{tikzpicture} \end{tabular}  \\
$C_{N}$  & $\Node{}{k}\Rightarrow\underbrace{ \node{}{k}-\cdots-\node{}{k}}_{N-1~\text{nodes}}\Leftarrow\node{}{k} \ \tikz[na]\node(C1){};$ &  \begin{tabular}[c]{@{}c@{}} \\ 
\begin{tikzpicture} [baseline=0, scale=0.9, transform shape]
\draw [ultra thick] (0,0)--(0,2.5) node[black,midway, xshift =-0.3cm, yshift=-1.5cm] {\footnotesize $\Ot^+$};
\draw (0.5,0)--(0.5,2.5); \draw (1,0)--(1,2.5); \draw (1.5,0)--(1.5,2.5); \draw (3,0)--(3,2.5); \draw (3.5,0)--(3.5,2.5); \draw (4,0)--(4,2.5); 
\draw [ultra thick] (4.5,0)--(4.5,2.5) node[black,midway, xshift =0.3cm, yshift=-1.5cm] {\footnotesize $\Ot^+$};
%
\draw [thick, color=red](0.02,1)--(0.5,1) node[black,midway, yshift=-0.2cm] {\scriptsize $k$} ;
%
\draw (0.5,0.9)--(1,0.9) node[black,midway, yshift=0.2cm] {\scriptsize $k$};
\draw (1,1.3)--(1.5,1.3) node[black,midway, yshift=0.2cm] {\scriptsize $k$} node[black,midway, yshift=-0.3cm] {\tiny D1} ;
\draw [loosely dotted] (1.5,1)--(3,1);
\draw (3,1.3)--(3.5,1.3) node[black,midway, yshift=0.2cm] {\scriptsize $k$} node[black,midway, xshift =0.3cm, yshift=1.4cm] {\footnotesize D3};
\draw (3.5,1.7)--(4,1.7) node[black,midway, yshift=0.2cm] {\scriptsize $k$};
%
\draw [thick, color=red](4,0.75)--(4.48,0.75) node[black,midway, yshift=0.2cm] {\scriptsize $k$};
%
\draw [decorate, decoration={brace, mirror}](0.5,-0.1)--(4,-0.1) node[black,midway,yshift=-0.5cm] {\footnotesize $N-1~\text{intervals}$};
\end{tikzpicture}  \end{tabular} \\
$D_{N}$ & $\node{}{k}-\node{\ver{}{k}}{2k}-\underbrace{\node{}{2k}-\cdots-\node{}{2k}}_{N-5~\text{nodes}}-\node{\ver{}{k}}{2k}-\Node{}{k}$ & \begin{tabular}[c]{@{}c@{}} \\  \begin{tikzpicture} [baseline=0, scale=0.9, transform shape]
\draw [ultra thick] (0,0)--(0,2.5) node[black,midway, xshift =-0.3cm, yshift=-1.5cm] {\footnotesize $\Ot^-$};
\draw (0.5,0)--(0.5,2.5); \draw (1,0)--(1,2.5); \draw (1.5,0)--(1.5,2.5); \draw (3,0)--(3,2.5); \draw (3.5,0)--(3.5,2.5); \draw (4,0)--(4,2.5); 
\draw [ultra thick] (4.5,0)--(4.5,2.5) node[black,midway, xshift =0.3cm, yshift=-1.5cm] {\footnotesize $\Ot^-$};
%
\draw [thick, color=red, rounded corners=0.75cm](0.5,1)--(-0.4,0.75)--(1,0.75) node[black,midway, xshift=-0.0cm, yshift=-0.2cm] {\scriptsize $k$} ;
%
\draw (0.5,0.9)--(1,0.9) node[black,midway, yshift=0.4cm] {\scriptsize $k$};
\draw (1,1.3)--(1.5,1.3) node[black,midway, yshift=0.4cm] {\scriptsize $2k$} ;
\draw (1,1.5)--(1.5,1.5) node[black,midway, yshift=-0.4cm] {\tiny D1} ;
\draw [loosely dotted] (1.5,1)--(3,1);
\draw (3,1.3)--(3.5,1.3) node[black,midway, yshift=0.4cm] {\scriptsize $2k$};
\draw (3,1.5)--(3.5,1.5) node[black,midway, xshift =0.3cm, yshift=1.4cm] {\footnotesize D3} ;
\draw (3.5,1.7)--(4,1.7) node[black,midway, yshift=0.5cm] {\scriptsize $k$};
%
\draw [thick, color=red, rounded corners=0.75cm](3.5,0.75)--(4.9,0.75) --(4,1) node[black,midway, xshift=-0.2cm, yshift=0.3cm] {\scriptsize $k$};
%
\draw [decorate, decoration={brace, mirror}](1,-0.1)--(3.5,-0.1) node[black,midway,yshift=-0.5cm] {\footnotesize $N-3~\text{intervals}$};
\end{tikzpicture}  \end{tabular} \\
\hline
\end{longtable}

It is worth noting the relation between the ``boundary'' of the Dynkin diagram and the type of the orientifold planes \cite{Hanany:2001iy}.  In particular,
\bi
\item the bifurcation corresponds to $\Ot^-$;
\item the double arrow going into the main body of the quiver corresponds to $\Ot^+$; and
\item the double arrow going out of the main body of the quiver corresponds to $\Ott^-$.
\ei

For the twisted case, we only focus on the twisted affine Lie algebras $A^{(2)}_{2N-1}$, $A^{(2)}_{2N}$, and $D^{(2)}_{N+1}$.  Their Dynkin diagrams can be realised on the worldvolume of the D1 branes in the following brane set-up \cite{Hanany:2001iy}.  Observe that the type of the orientifold threeplanes are in accordance with the rules stated above.
\begin{longtable}{|c|c|c|}
\hline
$g$  & Twisted affine Dynkin diagram of $g^{(2)}$ & Brane set-up\\
\hline
$A_{2N-1}$ & $\node{}{k}-\node{\blver{}{k}}{2k}-\underbrace{\node{}{2k}-\cdots-\node{}{2k}}_{N-4~\text{nodes}}-\node{}{2k} \Leftarrow \node{}{k}$ &
\begin{tabular}[c]{@{}c@{}} \\   
\begin{tikzpicture} [baseline=0, scale=0.9, transform shape]
\draw [ultra thick] (0,0)--(0,2.5) node[black,midway, xshift =-0.3cm, yshift=-1.5cm] {\footnotesize $\Ot^-$}; 
\draw (0.5,0)--(0.5,2.5); \draw (1,0)--(1,2.5); \draw (1.5,0)--(1.5,2.5); \draw (3,0)--(3,2.5); \draw (3.5,0)--(3.5,2.5); \draw (4,0)--(4,2.5); 
\draw [ultra thick] (4.5,0)--(4.5,2.5) node[black,midway, xshift =0.3cm, yshift=-1.5cm] {\footnotesize $\Ot^+$};
%
\draw [thick, color=red, rounded corners=0.75cm](0.5,1)--(-0.4,0.75)--(1,0.75) node[black,midway, xshift=-0.0cm, yshift=-0.2cm] {\scriptsize $k$} ;
%
\draw (0.5,0.9)--(1,0.9) node[black,midway, yshift=0.4cm] {\scriptsize $k$};
\draw (1,1.3)--(1.5,1.3) node[black,midway, yshift=0.4cm] {\scriptsize $2k$} ;
\draw (1,1.5)--(1.5,1.5) node[black,midway, yshift=-0.4cm] {\tiny D1} ;
\draw [loosely dotted] (1.5,1)--(3,1);
\draw (3,1.3)--(3.5,1.3) node[black,midway, yshift=0.4cm] {\scriptsize $2k$};
\draw (3,1.5)--(3.5,1.5) node[black,midway, xshift =0.3cm, yshift=1.4cm] {\footnotesize D3} ;
\draw (3.5,1.7)--(4,1.7) node[black,midway, yshift=0.5cm] {\scriptsize $2k$};
\draw (3.5,1.9)--(4,1.9);
%
\draw [thick, color=red](4,0.75)--(4.48,0.75) node[black,midway, yshift=0.2cm] {\scriptsize $k$};
%
%
\draw [decorate, decoration={brace, mirror}](1,-0.1)--(4,-0.1) node[black,midway,yshift=-0.5cm] {\footnotesize $N-2~\text{intervals}$};
\end{tikzpicture}
\end{tabular}
\\
$A_{2N}$    & $\Node{}{2k}\Leftarrow\underbrace{ \node{}{2k}-\cdots-\node{}{2k}}_{N-1~\text{nodes}}\Leftarrow\node{}{k}$  & 
\begin{tabular}[c]{@{}c@{}} \\ 
\begin{tikzpicture} [baseline=0, scale=0.9, transform shape]
\draw [ultra thick] (0,0)--(0,2.5) node[black,midway, xshift =-0.3cm, yshift=-1.5cm] {\footnotesize $\Ott^-$};
\draw (0.5,0)--(0.5,2.5); \draw (1,0)--(1,2.5); \draw (1.5,0)--(1.5,2.5); \draw (3,0)--(3,2.5); \draw (3.5,0)--(3.5,2.5); \draw (4,0)--(4,2.5); 
\draw [ultra thick] (4.5,0)--(4.5,2.5) node[black,midway, xshift =0.3cm, yshift=-1.5cm] {\footnotesize $\Ot^+$};
%
\draw [thick, color=red, rounded corners=0.75cm] (0.5,1)--(-0.4,0.75)--(0.5,0.75)  node[black,midway, xshift=0.2cm, yshift=0.4cm] {\scriptsize $2k$};
%
\draw (0.5,0.9)--(1,0.9) node[black,midway, yshift=0.2cm] {\scriptsize $2k$};
\draw (1,1.3)--(1.5,1.3) node[black,midway, yshift=0.2cm] {\scriptsize $2k$} node[black,midway, yshift=-0.3cm] {\tiny D1} ;
\draw [loosely dotted] (1.5,1)--(3,1);
\draw (3,1.3)--(3.5,1.3) node[black,midway, yshift=0.2cm] {\scriptsize $2k$} node[black,midway, xshift =0.3cm, yshift=1.4cm] {\footnotesize D3};
\draw (3.5,1.7)--(4,1.7) node[black,midway, yshift=0.2cm] {\scriptsize $2k$};
%
\draw [thick, color=red](4,0.75)--(4.48,0.75) node[black,midway, yshift=0.2cm] {\scriptsize $k$};
%
\draw [decorate, decoration={brace, mirror}](0.5,-0.1)--(4,-0.1) node[black,midway,yshift=-0.5cm] {\footnotesize $N-1~\text{intervals}$};
\end{tikzpicture}  \end{tabular}
\\
$D_{N+1}$  & $\Node{}{k}\Leftarrow\underbrace{ \node{}{k}-\cdots-\node{}{k}}_{N-1~\text{nodes}}\Rightarrow\node{}{k}$  & 
\begin{tabular}[c]{@{}c@{}} \\ 
\begin{tikzpicture} [baseline=0, scale=0.9, transform shape]
\draw [ultra thick] (0,0)--(0,2.5) node[black,midway, xshift =-0.3cm, yshift=-1.5cm] {\footnotesize $\Ott^-$};
\draw (0.5,0)--(0.5,2.5); \draw (1,0)--(1,2.5); \draw (1.5,0)--(1.5,2.5); \draw (3,0)--(3,2.5); \draw (3.5,0)--(3.5,2.5); \draw (4,0)--(4,2.5); 
\draw [ultra thick] (4.5,0)--(4.5,2.5) node[black,midway, xshift =0.3cm, yshift=-1.5cm] {\footnotesize $\Ott^-$};
%
\draw [thick, color=red, rounded corners=0.75cm] (0,0.75)--(0.5,0.75)  node[black,midway, xshift=0cm, yshift=0.2cm] {\scriptsize $k$};
%
\draw (0.5,0.9)--(1,0.9) node[black,midway, yshift=0.2cm] {\scriptsize $k$};
\draw (1,1.3)--(1.5,1.3) node[black,midway, yshift=0.2cm] {\scriptsize $k$} node[black,midway, yshift=-0.3cm] {\tiny D1} ;
\draw [loosely dotted] (1.5,1)--(3,1);
\draw (3,1.3)--(3.5,1.3) node[black,midway, yshift=0.2cm] {\scriptsize $k$} node[black,midway, xshift =0.3cm, yshift=1.4cm] {\footnotesize D3};
\draw (3.5,1.7)--(4,1.7) node[black,midway, yshift=0.2cm] {\scriptsize $k$};
%
\draw [thick, color=red](4,0.75)--(4.48,0.75) node[black,midway, yshift=0.2cm] {\scriptsize $k$};
%
\draw [decorate, decoration={brace, mirror}](0.5,-0.1)--(4,-0.1) node[black,midway,yshift=-0.5cm] {\footnotesize $N-1~\text{intervals}$};
\end{tikzpicture}  \end{tabular}
\\
\hline
\end{longtable}
As pointed out in \cite{Hanany:2001iy}, in the twisted affine cases, we need to turn on the Wilson line in the compact direction of the worldvolume of the D4 branes in order to make the algebra $\mathcal{G}_0$ enhance to the full twisted Lie algebra $g^{(L)}$.

\section{Dualities for symplectic gauge groups}
\label{sp}

In this section we discuss the 3d dualities 
for $USp(2N_c)$ SQCD obtained by compactifications
of 4d theories on $S^1$ in the presence of orientifolds.
We study both the affine and the twisted affine configurations.
We recover the various models discussed in the literature, namely
Aharony duality and dualities with linear and quadratic monopole superpotentials.

In order to fix the notations and the geometric setup we consider 
in 4d an NS brane extended along $x_{0,1,2,3,4,5}$
and an NS' brane extended along $x_{0,1,2,3,8,9}$.
The two fivebranes are separated along $x_6$. They are connected along this directions by a stack of D4 branes. These last are finite 
along $x_6$ and they fill  $x_{0,1,2,3}$. The flavor is obtained by adding to the picture a stack of D6 branes extended along 
$x_{0,1,2,3,7,8,9}$.
T-duality is performed along $x_3$, and in this way the NS branes are
compact along $x_3$ while the D4 and D6 branes become 
D3 and D5 respectively, and they are not extended anymore along $x_3$.
On this picture we can add a pair of orientifolds, as discussed above.

Our brane description distinguishes three possible 3d $\mathcal{N}=2$ gauge theories with symplectic gauge group
and fundamental matter.
They are summarized in the table below.
Let us study these three cases separately.
\begin{center}
\begin{longtable}{|c|c|c|c|c|}
\hline
$G_{\mathrm{ele}}$ & $G_{\mathrm{mag}}$ & $W_{\mathrm{ele}}$ & $W_{\mathrm{mag}}$ &Orientifolds\\
\hline
$USp(2N_c)$ & $USp(2N_f-2N_c-2)$    &  $W = 0$                       & $W = M q \tilde q + y Y$ 
&
\begin{tabular}{c}
\begin{tikzpicture}
\tikzstyle{every node}=[font=\scriptsize, node distance=0.45cm]
\tikzset{decoration={snake,amplitude=.4mm,segment length=2mm,
                       post length=0mm,pre length=0mm}}
\draw (0,0) circle (0.8cm);
\node[draw=none] at (-0.8,0) {{\red $\bigstar$}} node at (-1.2,0) {$\Ot^-$};
\node[draw=none] at (0.8,0) {{\red $\bigstar$}} node at (1.2,0) {$\Ot^+$};
\end{tikzpicture}
\end{tabular}
\\
\hline
$USp(2N_c)$ & $USp(2N_f-2N_c-4)$    &  $W = Y$  & $W = M q \tilde q +  y$ 
&
\begin{tabular}{c}
\begin{tikzpicture}[baseline]
\tikzstyle{every node}=[font=\scriptsize, node distance=0.45cm]
\tikzset{decoration={snake,amplitude=.4mm,segment length=2mm,
                       post length=0mm,pre length=0mm}}
\draw (0,0) circle (0.8cm);
\node[draw=none] at (-0.8,0) {{\red $\bigstar$}} node at (-1.2,0) {$\Ot^+$};
\node[draw=none] at (0.8,0) {{\red $\bigstar$}} node at (1.2,0) {$\Ot^+$};
\end{tikzpicture}
\end{tabular}
\\
\hline
$USp(2N_c)$ & $USp(2N_f-2N_c-2)$    &  $W = Y^2$  & $W = M q \tilde q +  y^2$ 
&
\begin{tabular}{c}
\begin{tikzpicture}[baseline]
\tikzstyle{every node}=[font=\scriptsize, node distance=0.45cm]
\tikzset{decoration={snake,amplitude=.4mm,segment length=2mm,
                       post length=0mm,pre length=0mm}}
\draw (0,0) circle (0.8cm);
\node[draw=none] at (-0.8,0) {{\red $\bigstar$}} node at (-1.2,0) {$\Ott^-$};
\node[draw=none] at (0.8,0) {{\red $\bigstar$}} node at (1.2,0) {$\Ot^+$};
\end{tikzpicture}
\end{tabular}
\\
\hline
\end{longtable}
\end{center}
%
%
%
%
%
%
%
\subsection{$USp(2N_c)$ with $W=0$}
\label{USp1}
%
%
%
This cases corresponds to the original Aharony duality discussed in \cite{Aharony:1997gp}.
In the brane picture it corresponds to the setup with an $\Ot^+$ plane at $x_3=0$ and 
an $\Ot^-$ at $x_3= \pi R$.
Such an orientifold boundary condition corresponds to the the twisted affine algebra $A_{2N_c-1}^{(2)}$:
\begin{equation}
\node{}{}-\node{\ver{}{}}{}-\underbrace{\node{}{}-\node{}{}-\cdots-\node{}{}-\node{}{}}_{\text{$(N_c-3)$ nodes}}\Leftarrow\node{}{}~.
\end{equation}
Observe that there is also a configuration with the two 
orientifolds exchanged.
Such a configuration corresponds to turning on opportune Wilson 
lines in the 4d setup.

In this brane setup we consider $N_c$ D3 and $N_f$ D5 on top of $\Ot^+$ at $x_3=0$
while we do not have any further brane at $x_3= \pi R$.  This gives rise to the $USp(2N_c)$ gauge theory with $2N_f$ fundamentals.

Let us now discuss the brane configuration after the transition through infinite coupling.  At $x_3=0$, where $\Ot^+$ is located, we have $N_f-N_c-1$ physical D3 branes, where $-1$ is there to cancel the charge of the $\Ot^+$ plane.  On the other hand, at $x_3= \pi R$, where $\Ot^-$ is located, we have one physical D3 brane there to cancel the charge of $\Ot^-$.
This configuration gives rise to a $USp(2(N_f-N_c-1))$ gauge theory with $2N_f$ chirals at $x_3=0$, and a pure $SO(2)$ gauge sector at  $x_3= \pi R$.  There is also an  interaction
\footnote{
A monopole $Y$ is here identified with a Coulomb branch coordinate 
$\Sigma$. On the Coulomb branch the latter are holomorphic combinations $\Sigma \equiv i \varphi + \frac{\sigma}{g^2}$ of the 
dual photon $\varphi$ and of the real scalar $\sigma$ in the vector multiplet, where  $g$ represents the gauge coupling.
}
\begin{equation}
\label{Waha}
W = \mathe^{ \Sigma-\Sigma_{1}}
\end{equation}
between the monopole $y=\mathe^{- \Sigma_{1}}$ of the dual $USp(2(N_f-N_c-1))$
gauge group and the monopole $Y = \mathe^{ \Sigma}$ of the $SO(2)$ sector.

Observe that in this case, in absence of D-branes in the electric sector we can consider the large 
T-dual radius limit, describing a pure 3d gauge theory, On the magnetic side we can dualize the $SO(2)$ 
gauge sector, and indeed we remain with the singlet $Y$, acting with the superpotential (\ref{Waha}), corresponding to 
the interaction $W = y Y$ of the Aharony duality.
%
%
%
\subsection{$USp(2N_c)$ with $W=Y$}
%
%
%
This case corresponds to the affine $C^{(1)}_{N_c}$ case, corresponding to the circle compactification of the 4d $USp(2N_c)$ theory 
and of its Intriligator-Pouliot dual description, whose affine Dynkin diagram is
\be
\node{}{}\Rightarrow\underbrace{ \node{}{}-\cdots-\node{}{}}_{N_c-1~\text{nodes}}\Leftarrow\node{}{}
\ee

In terms of branes, we have an $\Ot^+$ plane at $x_3=0$ and 
an $\Ot^+$ at $x_3= \pi R$.
In this brane setup we consider $N_c$ D3 and $N_f-1$ D5 at $x_3=0$ (with their images) 
while we have one D5  $x_3= \pi R$ (with its image).  This system gives rise to the $USp(2N_c)$ gauge theory with $2N_f$ fundamentals with $W= Y$, where the monopole superpotential can be read from the brane configuration at $x_3= \pi R$.

Let us now discuss the brane configuration after the transition through infinite coupling.  At $x_3=0$, where one of the $\Ot^+$ planes is located, we have $(N_f-1)-N_c-1$ physical D3 branes, where the last $-1$ is there to cancel the charge of $\Ot^+$.  This gives rise to a $USp(2(N_f-N_c-2))$ gauge theory with $2N_f$ fundamentals. At $x_3= \pi R$, where the other $\Ot^+$ is located, we have $(1-0)-1=0$ D3 brane, where $1$ and $0$ denote numbers of D5 and D3 branes before the transition and the last $-1$ is there to cancel the charge of $\Ot^+$.
The absence of D3 branes at $x_3= \pi R$ for both the electric and the magnetic descriptions
allow us to place the extra D5 branes from this position to $x_3=0$.
This system gives rise to the dual theory, which is the $USp(2(N_f-N_c-2))$ gauge theory with $2N_f$ fundamentals, singlets $M$ and $W=  M q q + y$, where the monopole superpotential can be read from the brane configuration at $x_3= \pi R$.
%
%
%
\subsection{$USp(2N_c)$ with $W=Y^2$}
\label{USp2}
%
%
%
This case corresponds to another twisted affine compactification.
At the geometric level we have an $\Ot^+$ plane at $x_3=0$ and 
an $\Ott^-$ at $x_3= \pi R$.
Such an orientifold boundary condition corresponds to the the twisted affine algebra $A_{2N_c}^{(2)}$:
\begin{equation}
\node{}{} \Leftarrow \underbrace{\node{}{}-\node{}{}- \cdots -\node{}{}- \node{}{}}_{\text{$(N_c-1)$ nodes}} \Leftarrow \node{}{}~.
\end{equation}
Observe that there is also a configuration with the two 
orientifolds exchanged.
Such a configuration corresponds to turning on opportune Wilson 
lines in the 4d setup.

In this brane setup we consider $N_c$ D3 and $N_f$ D5 (with their images), along with $\Ot^+$, at $x_3=0$ ,
while we have a half physical D3 brane stuck on the $\Ott^-$ plane at $x_3= \pi R$. This gives rise to the $USp(2N_c)$ gauge theory with $2N_f$ fundamentals and $W=Y^2$, where the monopole superpotential can be read off from the configuration at $x_3= \pi R$.

After the transition through infinite coupling we are left with $N_f-N_c-1$ D3 branes at
$x_3=0$, where $-1$ is there to cancel the charge of $\Ot^+$.  There is a half physical D3 brane stuck on $\Ott^-$ at $x_3= \pi R$.
This system gives rise to the dual theory, namely the $USp(2(N_f-N_c-1))$ gauge theory with $2N_f$ fundamentals, singlets $M$ and
the superpotential $W = M q q +  y^2$.

\section{Dualities for unitary gauge groups}
\label{uni}

In this section we study unitary gauge groups, corresponding to 
placing $N_c$ D3 branes and $N_f$ D5 branes at 
$x_3 = \tfrac{\pi}{2} R$.
Depending on the choice of the orientifolds we also place
other D3 or D5 branes at $x_3=0$ and $x_3= \pi R$.
By exchanging the position of the NS branes we 
generate the dual description.
We can summarize the results of this section in the following table.
\begin{center}
\begin{longtable}{|c|c|c|c|c|}
\hline
$G_{\mathrm{ele}}$ & $G_{\mathrm{mag}}$ & $W_{\mathrm{ele}}$ & $W_{\mathrm{mag}}$ &Orientifolds\\
\hline
$U(N_c)$ & $U(N_f-N_c)$    &  $W = 0$                       & $W = M q \tilde q + T_+ t_+ + T_- t_-$ 
&
\begin{tabular}{c}
\includegraphics[width=2cm]{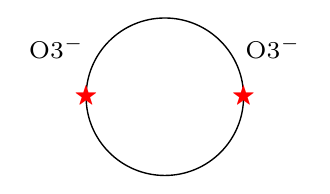}
\end{tabular}
\\
\hline
$U(N_c)$ & $U(N_f-N_c-2)$ &  $W = T_+ + T_-$         & $W = M q \tilde q +  t_+ + t_-$ 
&
\begin{tabular}{c} 
\includegraphics[width=2cm]{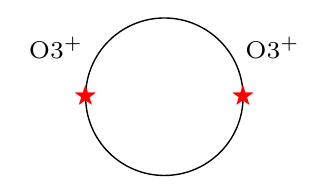}
\end{tabular} 
\\
\hline
$U(N_c)$ & $U(N_f-N_c-1)$ &  $W = T_+$                   & $W = M q \tilde q +  t_- + T_- t_+$ 
&
\begin{tabular}{c} 
\includegraphics[width=2cm]{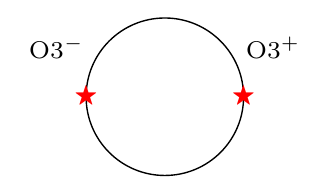}
\end{tabular} 
\\
\hline
$U(N_c)$ & $U(N_f-N_c)$    &  $W = T_+^2 + T_-^2$  & $W = M q \tilde q +  t_+^2 +  t_-^2$ 
&
\begin{tabular}{c} 
\includegraphics[width=2cm]{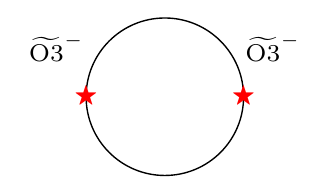}
\end{tabular} 
\\
\hline
$U(N_c)$ & $U(N_f-N_c)$    &  $W =T_+^2 $               & $W = M q \tilde q + t_-^2 + T_- t_+$ 
&
\begin{tabular}{c} 
\includegraphics[width=2cm]{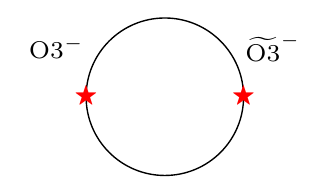}
\end{tabular} 
\\
\hline
$U(N_c)$ & $U(N_f-N_c-1)$ &  $W = T_+^2 + T_-$     & $W = M q \tilde q +  t_-^2 +  t_+$ 
&
\begin{tabular}{c} 
\includegraphics[width=2cm]{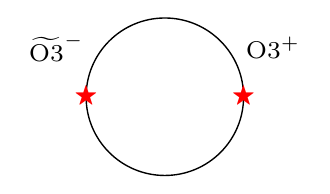}
\end{tabular} 
\\
\hline
\end{longtable}
\end{center}
As already anticipated in the introduction, most of the models have already been discussed in the literature.
However there is a new case, so far overlooked, corresponding to 
$U(N_c)$ SQCD with $W = T_+^2 + T_-$.
Observe that a full classification should have nine inequivalent cases.
The other three cases that we did not discuss here correspond to the pairs $(\Ot^-,\Ott^-)$, $(\Ot^-,\Ot^+)$ and $(\Ot^+,\Ott^-)$.
These cases can be obtained by 
by charge conjugation on
$(\Ott^-,\Ot^-)$, $(\Ot^+,\Ot^-)$ and $(\Ott^+,\Ot^+)$
respectively.

In the following we discuss the various cases separately, showing how to construct the 3d dualities from the brane picture in each case.

\subsection{$U(N_c)$ with $W=0$: Aharony duality}

Aharony duality can be constructed by reducing a 4d $SO(2N)$ gauge theory with $N_f$ flavors
on $S^1$ and considering the vacuum corresponding to $N_c$ D3 and $N_f$ D5 
at $x_3 = \tfrac{\pi}{2} R$.  The corresponding 4d theory on $S^1$ has a monopole superpotential $W=\eta Z$, where $Z$ is the KK monopole operator\footnote{Semiclassically, this corresponds to $Z \sim 
\mathe^{\Sigma_1+\Sigma_2}$ in the notation of \cite{Aharony:2013kma}.} \cite{Aharony:2013kma}.

Since we do not have any extra D3 or D5 at $x_3=0,\pi R$,  it signals
the fact that we can sent the radius and the monopole superpotential to zero.  This is in agreement with the discussion below (5.2) of \cite{Aharony:2013kma}. The resulting theory is thus a 3d $U(N_c)$ gauge theory with $N_f$ flavors and zero superpotential.

In the dual picture we have $N_f-N_c$ D3 and $N_f$ D5 at 
$x_3 = \tfrac{\pi}{2} R$. Moreover we have a 
D3 and its image at $x_3=0$ and at $x_3= \pi R$. In this case we have to dualize these $SO(2)  = U(1) $ gauge theories.
In the dual picture the monopole corresponds to a singlet and it can be identified with the electric monopole
acting as a singlet in this dual phase. This is compatible with claim that this brane picture
represents the dual phase of Aharony duality.

\subsection{$U(N_c)$ with $W=T_++T_-$}

This duality has been already studied in \cite{Benini:2017dud}, and it corresponds to the reduction 
of a 4d $USp(2N_c)$ SQCD with $2 N_f$ fundamentals.  Upon putting this theory on $S^1$, a superpotential $W=\eta Y$ is generated.   
In the electric theory one needs to consider a vacuum with
$N_c$ D3 and $N_f-2$ D5 at $x_3 = \tfrac{\pi}{2} R$.
Moreover there is a pair of one D5 brane and its image at both $x_3=0$ and $x_3= \pi R$.  

The dual picture has $N_f-N_c-2$ D3 and $N_f-2$ D5 at $x_3 = \tfrac{\pi}{2} R$.
and again a pair of one D5 brane and its image at both $x_3=0$ and $x_3= \pi R$.
The absence of $D3$ branes at $x_3=0$ and $x_3= \pi R$
in both phases allows to re-collect all the D5 at 
$x_3 = \tfrac{\pi}{2} R$ in both phases.
Furthermore the monopole superpotential can be read from the
spectrum of D1 branes connecting the stack of D3 branes at the orientifolds.

\subsection{$U(N_c)$ with $W=T_+$}
We start by reducing the 4d $USp(2N_c)$ gauge theory with $2N_f$ fundamentals and its dual on a circle.
The brane system consists of an $\Ot^+$ plane at $x_3=0$ and an $\Ot^-$ plane at
$x_3= \pi R$.  The electric theory on $S^1$ has a superpotential $W =\eta Y$ and the dual theory has gauge group $USp(2N_f-2N_c-4)$.
Such an orientifold boundary condition corresponds to the the twisted affine algebra $A_{2N_c-1}^{(2)}$:
\be
\node{}{}-\node{\ver{}{}}{}-\underbrace{\node{}{}-\node{}{}-\cdots-\node{}{}-\node{}{}}_{\text{$(N_c-3)$ nodes}}\Leftarrow\node{}{}~.
\ee
For our aims the configuration with the two 
orientifolds exchanged is completely equivalent, and it 
corresponds to turning on opportune Wilson 
lines in the 4d setup.

In the electric theory we have $N_c$ D3 and $N_f-1$ D5 at 
$x_3 =\tfrac{\pi}{2} R$. 
We consider also one D5 brane and its image at
$x_3=0$ while we do not have any D-brane at $x_3= \pi R$.

In the dual model we have $N_f-N_c-1$ D3 and $N_f-1$ D5 at 
$x_3 =\tfrac{\pi}{2} R$, along with one D5 and its image at $x_3=0$ and 
one D3 and its image at $x_3= \pi R$.
We are free to connect all the D5 at 
$x_3 = \tfrac{\pi}{2} R$ in both phases and to dualize the $SO(2)$ gauge theory 
into a scalar.
The final duality relates a $U(N_c)$ gauge theory with $N_f$ pairs of fundamentals and antifundamentals
with monopole superpotential $W = T_+$ with a $U(N_f-N_c-1)$ gauge theory with $N_f$ pairs of dual 
fundamentals and antifundamentals, with superpotential $W = M q \tilde q + t_-+ t_+ T_-$ where $M$ corresponds
to the meson of the electric theory and $T_-$ is the dual photon of the $SO(2)$ gauge theory and has the same quantum
numbers of the anti-monopole of the electric theory.

Let us end this subsection by mentioning a puzzle regarding the twisted compactification in this case.  As we mentioned at the beginning, we start from the $USp(2N_c)$ gauge theory on $S^1$.  There are two options to obtain such a gauge algebra from 4d, namely
\ben
\item  $A_{2 \ell} = su(2\ell+1) \rightarrow C_\ell = usp(2 \ell)$; or
\item  $D_{\ell+1} = so(2\ell+2) \rightarrow C_\ell = usp(2 \ell)$
\een
For option 1, the 4d Seiberg duality between an $SU(2N_c+1)$ gauge theory with $2N_f$ flavors and an $SU(2N_f-2N_c-1)$ gauge theory with $2N_f$ flavors becomes a duality between a $usp(2N_c)$ gauge theory and a $USp(2N_f-2N_c-2)$ gauge theory; however, the latter is not $USp(2N_f-2N_c-4)$ as expected.  For option 2, the 4d Seiberg duality between an $SU(2N_c+2)$ gauge theory with $2N_f$ flavors and an $SU(2N_f-2N_c-2+4)=SU(2N_f-2N_c+2)$ gauge theory with $2N_f$ flavors becomes a duality between a $USp(2N_c)$ gauge theory and a $USp(2N_f-2N_c)$ gauge theory; however, the latter is not $USp(2N_f-2N_c-4)$ as expected.   One possibility to resolve this puzzle is that in this brane set up there is a Wilson line that could break the $USp(2N_f-2N_c-2)$ gauge group to the $USp(2N_f-2N_c-4)$ gauge group (or from the $USp(2N_f-2N_c)$ gauge group to the $USp(2N_f-2N_c-4)$ gauge group).  We leave this for future work.

\subsection{$U(N_c)$ with $W=T_+^2 + T_-^2$}
The 3d duality in this case can be realised by starting from the following 4d theories on $S^1$ with a special orthogonal gauge algebra. 
The latter can be obtained such a duality from 4d Seiberg duality by twisted compactification as follows.  Let us use the non-trivial outer-automorphism of the $A_{2\ell-1} = su(2\ell)$ algebra to twist and obtain the $B_\ell=so(2\ell+1)$ algebra:
\begin{figure}[h]
\def\circn#1{\tikz[na]\node(#1){$\circ$};}
\def\seg{\,\tikz[na] \draw(0,0)--(1em,0) ;}
\def\ud{{\color{red}\updownarrow}}
\[
\begin{array}{l c@{}c@{}c@{}c@{}c@{\hskip-.3em}c@{}c}
&\circ &  \seg & \circ & \cdots  &\circn{A}   \\
 A_{2\ell- 1}~~~&\ud&&\ud && \ud & &\circn{B} \\
\qquad &\circ & \seg  & \circ  & \cdots  & \circn{C} &  \\
B_\ell ~~~ & \circ & \seg  &\circ &  \cdots  & \circ & \Rightarrow  & \circn{} \\
\end{array}
\]

\begin{tikzpicture}[overlay,thin]
        \draw (A) --  (B) --(C);
\end{tikzpicture}
\end{figure}
The Seiberg duality between the $SU(2\ell)$ gauge theory with $2N_f$ flavors and the $SU(2N_f -2\ell)$ gauge theory with $2N_f$ flavors and singlets then becomes a duality between a theory with the $so(2\ell+1)$ gauge algebra and a theory with the $so(2N_f-2\ell+1)$ gauge algebra after twisting.  In this paper, we shall not go into any further detail of the duality between theories with the orthogonal gauge algebra.

The brane system of such theories with the $B$-type orthogonal gauge algebras contains a pair of $\Ott^-$ planes, one at $x_3=0$ and the other at
$x_3= \pi R$. Recall that on each $\Ott^-$ plane, there is a half D3 brane stuck there.  The orientifold boundary condition corresponds to the twisted affine algebra $D_{N_c+1}^{(2)}$:
\be
\underbrace{\node{}{} \Leftarrow \node{}{} -\node{}{}-\cdots -\node{}{}-\node{}{} \Rightarrow \node{}{}}_{\text{($N_c+1$) nodes}}~.
\ee

In the electric theory we have $N_c$ D3 and $N_f$ D5 at 
$x_3 =\tfrac{\pi}{2} R$: the gauge theory corresponds
to $U(N_c)$ with $N_f$ pairs of fundamentals and anti-fundamentals and superpotential
$W = T_+^2 + T_-^2$ corresponding to the spectrum of D1 branes connecting the D3 branes at the orientifold and the 
D3 branes on the stack.

In the dual configuration we have $N_f-N_c$ D3 and $N_f$ D5 at 
$x_3 =\tfrac{\pi}{2} R$,
a single D3 at $x_3=0$ and another at $x_3= \pi R$.
The gauge theory corresponds to $U(N_f-N_c)$ with $N_f$ pairs of fundamentals and anti-fundamentals and superpotential
$W = M q \tilde q +  t_+^2 + t_-^2$.

\subsection{$U(N_c)$ with $W=T_+^2$}

This duality can be constructed by reducing a 4d model with  $so(2N_c+1)$ gauge algebra and $2 N_f$
vectors on $S^1$.  The brane set-up contains an $\Ott^-$ at $x_3=0$ where there is a half D3 brane stuck there, 
together with a $\Ot^-$ at $x_3=\pi R$.
We consider the vacuum corresponding to $N_c$ D3 and $N_f$ D5 at $x_3 = \tfrac{\pi}{2} R$.
This electric theory corresponds to a $U(N_c)$ model with $N_f$ pairs of
fundamentals and antifundamentals and superpotential $W = T_+$.
As discussed above we consider a decoupling limit without the generation of any
monopole superpotential arising from the $\Ot^-$ plane.

The dual model is obtained by exchanging the position of the NS branes and it corresponds to 
consider $N_f-N_c$ D3 branes and $N_f$ D5 branes at 
$x_3 = \tfrac{\pi}{2} R$ and again a half $D3$ stuck on $\Ott^-$ at 
$x_3=0$. Furthermore there are one D3 brane and its image at $x_3= \pi R$ on the $\Ot^-$ plane.
The total amount of D3 branes in this setup corresponds to the total amount of D4
in the 4d theory as it should be. Indeed the dual 4d model corresponds to a theory with algebra 
$so(2N_f-2N_c+3)$; this is because we can recollect all $N_f-N_c+1$ D3 branes on the $\Ott^-$ plane.
The $SO(2)$ gauge theory at $x_3= \pi R$ can be dualized to a scalar, and this scalar corresponds to the 
electric monopole acting as a singlet in the dual phase.
All in all the dual model corresponds to a $U(N_f-N_c)$ gauge theory with 
with $N_f$ pairs of
fundamentals and antifundamentals and superpotential $W = t_-^{2}+T_- t_+$,
being $T_-$ the singlet obtained by dualizing the $SO(2)$ gauge theory.

\subsection{$U(N_c)$ with $W=T_+^2+T_-$}

The 3d duality in this case can be realised by starting from the following 4d theories on $S^1$ with a symplectic gauge algebra and a quadratic monopole superpotential:
\be \label{UspTp2Tm}
\begin{split}
&\text{(A): $USp(2N_c)$ SQCD with $2N_f$ chirals and $W=Y^2$.} \\ 
&\text{(B): $USp(2N_f-2N_c-2)$ SQCD with $2N_f$ chirals,} \\
& \qquad  \text{singlets $M$ and $W= Mq q +\hat{Y}^2$.}
\end{split}
\ee
We may obtain such a duality from 4d Seiberg duality by twisted compactification as follows.  Let us use the non-trivial outer-automorphism of the $A_{2\ell} = su(2\ell+1)$ algebra to twist and obtain the $C_\ell=usp(2\ell)$ algebra.  The Seiberg duality between the $SU(2 N_c+1)$ gauge theory with $2N_f$ flavors and the $SU(2N_f -2N_c-1)$ gauge theory with $2N_f$ flavors and singlets then becomes a duality between the $USp(2N_c)$ gauge theory and the $USp(2N_f-2N_c-2)$ gauge theory after twisting.  These are indeed the gauge groups in \eref{UspTp2Tm}, as required.

The brane system contains an $\Ot^+$ 
at $x_3=0$ (or at $x_3= \pi R$) and a $\Ott^-$ at $x_3= \pi R$ (or at $x_3 = 0$).
In the presence of D3 branes this gives raise to the superpotential $W=T_+^2+T_-$ (or $W=T_-^2+T_+$).
This case is interesting because it has been overlooked so far in the literature, while it seems
a natural possibility to investigate in the brane setup. 
The orientifold boundary condition corresponds to the twisted affine algebra $A_{2N_c}^{(2)}$:
\be
\node{}{} \Leftarrow \underbrace{\node{}{}-\node{}{}- \cdots -\node{}{}- \node{}{}}_{\text{$(N_c-2)$ nodes}} \Leftarrow \node{}{}~.
\ee

For definiteness, let us fix $\Ot^+$ to be at $x_3=0$ and $\Ott^-$ to be at $x_3= \pi R$.
In this case, we put $N_c$ D3 and $N_f-1$  D5 at
$x_3 = \tfrac{\pi}{2} R$.
Moreover we have one D5 brane (and its image) at $x_3 = 0$, as well as a half physical D3 brane stuck on the $\Ott^-$ plane at $x_3= \pi R$.

In the dual configuration we have  $N_f-N_c-1$ D3 and $N_f-1$  D5 at $x_3= \pi R$.
We also have one D5 brane (and its image)  at $x_3 = 0$ and a half physical D3 brane stuck at $\Ott^-$ at $x_3= \pi R$.
We can furthermore re-connect the D5 brane at 
$x_3= \tfrac{\pi}{2} R$ and the final configuration 
represents a $U(N_f-N_c-1)$ gauge theory with 
$N_f$ pairs of fundamentals and anti-fundamentals and 
superpotential $W = M q \tilde q + t_-^2 + t_+$.

\subsubsection{A further argument: the $S^3$ partition function}
\label{further}
We can provide a further argument for the validity of the duality just proposed by studying the
3-sphere partition function.
We can indeed prove analytically the integral identity between the electric and the magnetic side.
The partition function for a $U(N_c)$ gauge theory 
with $N_f$ pairs of fundamentals can be read from formula (\ref{formgen}) in appendix \ref{AppA}, by setting $\tau = \omega$:
\begin{eqnarray}
Z_{U(N_c)}(\mu;\nu;\eta) \equiv  Z_{U(N_c)}(\mu;\nu;\omega;\eta)
\end{eqnarray}
At this point we can consider the duality between 
$U(N_c)$ with $N_f$ pairs of fundamentals and 
antifundamentals and superpotential $W = T_+$ and
$U(N_f-N_c-1)$ with $N_f$ pairs of fundamentals and 
antifundamentals and superpotential $W = M q \tilde q + t_- + t_+ T_- $.

The matching between the electric and the magnetic partition functions has been proven 
for this case by \cite{Benini:2017dud}.
The identity is
\begin{eqnarray}
 \label{eq:Aharony}
Z_{U(N_c)}(\mu;\nu;\eta-2\omega)
&=&
\mathe^{\frac{\mathi \pi}{2} \sum_{a=1}^{N_f} (\mu_a^2-\nu_a^2)}
\Gamma_h (\lambda) 
\prod_{a,b=1}^{N_f}
\Gamma_h(\mu_a + \nu_b)
\nonumber \\
&\times&
Z_{U(N_f-N_c-1)}(\omega-\mu;\omega-\nu;\eta)
\end{eqnarray}
where the parameters $\mu$, $\nu$ and $\eta$ are constrained
by
\begin{equation}
\label{BCAH}
 \sum_{a=1}^{N_f} (\mu_a +\nu_a) +\frac{\eta}{2} =  \omega(N_f-N_c)
\end{equation}
Following the mathematical literature, from now on, we will refer to this and similar types of identities between the parameters entering in the partition function as balancing conditions.
From \eref{eq:Aharony} one will now prove the identity for the case at hand.

This can be understood by a field theoretical analysis as follows:
deforming the electric side of the duality by adding a superpotential 
term proportional to $T_-^2$
we impose the balancing conditions
 \begin{equation}
 \label{eq:constraints}
\eta = \omega,
\quad
 \sum_{a=1}^{N_f} (\mu_a +\nu_a)  = 2 \omega
 \left(N_f-N_c-\frac{1}{2} \right)
\end{equation}
By plugging (\ref{eq:constraints}) in the identity (\ref{eq:Aharony})
and by using the fact that $\Gamma_h(\omega) = 1$ 
one arrives at the identity 
\begin{eqnarray}
 \label{eq:quadratic}
Z_{U(N_c)}(\mu;\nu;-\omega)
&=&
\mathe^{\frac{\mathi \pi}{2} \sum_{a=1}^{N_f} (\mu_a^2-\nu_a^2)}
\prod_{a,b=1}^{N_f}
\Gamma_h(\mu_a + \nu_b)
\nonumber \\
&\times&
Z_{U(N_f-N_c-1)}(\omega-\mu;\omega-\nu;\omega)
\end{eqnarray}
with the balancing conditions  (\ref{eq:constraints}), that provides the 
equality between the partition functions that we are looking for.
\\
\\
Furthermore the brane picture suggests an RG flow interpolating between 
$USp(2N_c)$ with $2N_f$ fundamentals and $W = Y^2$ and
$U(N_c)$ with $N_f$ pairs of fundamentals and anti-fundamentals and 
$W = T_+ + T_-^2$ (or $W = T_- + T_+^2$).
Moreover this flow should interpolate between the two dualities involving 
symplectic and unitary groups respectively.
Here we check these expectation against the partition function.
This provides a further argument in favor of the new duality 
for unitary theories and monopole superpotential  $W = T_+ + T_-^2$ (or $W = T_- + T_+^2$).

\subsection*{The electric flow}

The partition function for $USp(N_c)$ with $2N_f$ fundamentals
can be read from formula (\ref{formgen2}) in appendix \ref{AppA}
by setting $\tau = \omega$.
We have
\begin{equation}
Z_{USp(2N_c)}(m) \equiv Z_{USp(2N_c)}(m;\omega)
\end{equation}
On the electric side we then consider the partition function
$Z_{\mathrm{ele}} = Z_{USp(2N_c)}(m)$.
 The quadratic monopole superpotential imposes the balancing condition 
 \begin{equation}
 \sum_{a=1}^{2N_f} m_a = \omega(2N_f-2N_c-1)
  \end{equation} 
  We then consider the Higgs flow 
triggered by the shift
\footnote{
We could have chosen the opposite signs for $s$. This choice corresponds
to $W = T_+ + T_-^2$, while the opposite choice corresponds to 
$W = T_- + T_+^2$ }
\begin{equation}
\label{HF}
\sigma_a \rightarrow \sigma_a-s,
\quad
a = 1,\dots,N_c 
\end{equation}
and the real mass flow triggered by
\begin{equation}
\label{MF}
m_i \rightarrow \mu_i+s
\quad
m_{i+N_f} \rightarrow \nu_i-s
\quad
i= 1,\dots N_f
\end{equation}
By plugging (\ref{HF}) and (\ref{MF}) in $Z_{USp(2N_c)}(m)$
and by
computing the large $s$ limit using formula (\ref{intout}),
   we arrive at the partition function
of the $U(N_c)$ gauge theory with 
$N_f$ pairs of fundamentals and anti-fundamentals  and 
$W = T_+ + T_-^2$ 
 \begin{equation}
\label{PFe}
Z_{\mathrm{ele}}
=
\frac{\mathe^{\frac{\mathi \pi}{2}( As+B)}}{ N_c!}
Z_{U(N_c)} (\mu;\nu;-\omega)
 \end{equation}
where
\begin{eqnarray}
A = -4 N_c^2 \omega,
\quad
B = N_c \left(\sum _{i=1}^{N_f} \left(\mu _i^2-\nu _i^2\right)-2 \omega  \sum _{i=1}^{N_f} \left(\mu _i-\nu _i\right)\right)
\end{eqnarray}

\subsection*{The magnetic flow}

On the magnetic  side the partition function is 
\begin{equation}
\label{PF2}
Z_{\mathrm{mag}} = 
\prod_{1\leq a<b \leq 2 N_f }\Gamma_h(m_i + m_j)
Z_{USp(2\widetilde{N}_c)}( \widetilde{m}_i)
\end{equation}
with $\widetilde{N}_c = N_f-N_c-1$
and $\widetilde m_j = \omega-m_j$.
The dual Higgs flow is triggered by
\begin{equation}
\label{HF}
\sigma_a \rightarrow \sigma_a+s,
\quad
a = 1,\dots,\widetilde{N}_c 
\end{equation}
while the real mass flow can be read by using the duality map 
from the electric one.
In  the large $s$ limit we arrive at the partition function
of the $U(\widetilde{N}_c)$ gauge theory with 
$N_f$ pairs of fundamentals and anti-fundamentals, $N_f^2$ singlets
$M_{ij}$  (with $i,j = 1,\dots N_f$) and  $W = M q \tilde q + t_- + t_+^2$ 
 \begin{equation}
\label{PFm}
Z_{\mathrm{mag}}
=
\mathe^{\widetilde As + \widetilde B}
\prod_{i,j=1}^{N_f} \Gamma_h(\mu_i + \nu_j)
Z_{U(\widetilde{N}_c)} (\widetilde{\mu},\widetilde{\nu};\omega)
\end{equation}
where $\widetilde {\mu}= \omega - \mu$, 
$\widetilde{\nu} = \omega-\nu$
and
\begin{eqnarray}
\widetilde {A} = A ,
\quad
\widetilde {B} = 
(N_c-1) \sum _{i=1}^{N_f} (\mu _i^2-\nu _i^2)
-
\omega (2 N_c-1) \sum _{i=1}^{N_f} (\mu _i-\nu _i)
\end{eqnarray}
Moreover, using the fact that $\sum \mu_i = \sum \nu_i$
we can equate (\ref{PFe}) and (\ref{PFm}) and we are left with 
the identity (\ref{eq:quadratic}) as expected.

\section{Dualities with tensorial matter}
\label{plaw}

In this section we study 3d $\mathcal{N}=2$ dualities 
in the presence of tensorial matter fields and quadratic
monopole superpotential.
The analysis is inspired from the discussion in \cite[sec. 4.1.1]{Amariti:2018gdc} and here in (\ref{further}).
The idea consists in deforming an electric 
 duality by a quadratic monopole superpotential
and to find the dual deformation on the magnetic side.
These superpotentials impose a set of constraints
on the complex combinations of real masses and R-charges appearing 
as parameters in the partition function.
After fixing these constraints, the identities relating the partition 
functions of the parent dualities become new identity among the 
partition functions of the new dualities.
In this last step some singlets may disappear from the 
identity because they contribute as fields with holomorphic 
masses, i.e. their partition function is equal to one.

\subsection{$U(N_c)$ gauge group}
\label{subsecUadj}
We start our analysis by considering SQCD with $U(N_c)$ gauge groups, $N_f$ fundamentals $Q$ and 
antifundamentals $\widetilde Q$
and one adjoint matter field with superpotential

\begin{equation}
W_\mathrm{ele} = \tr X^{k+1}
\end{equation}
This theory is dual to $U(k N_f-N_c)$ SQCD with 
$N_f$ dual fundamentals $q$ and 
antifundamentals $\tilde q$, $k$ mesons $M_j = Q X^j \widetilde Q$,
$j=0,\dots,k-1$, an adjoint $Y$ with superpotential
\begin{equation}
W_\mathrm{mag} = \tr Y^{k+1} + \sum_{j=0}^{k-1}
M_j q Y^{k-1-j} \tilde q 
+
\sum_{j=0}^{k-1} 
(T_j t_{k-1-j}+ \widetilde T_{j} \widetilde t_{k-1-j})
\end{equation}
where $T_j= T_0 \tr X^j$,
$\widetilde T_j= \widetilde T_0 \tr X^j$,
$t_j= t_0 \tr Y^j$,
and 
$\widetilde t_j= \widetilde t_0 \tr Y^j$, and
$T_0$, $\widetilde T_0$, $t_0$ and $\widetilde t_0$
are the bare monopoles and anti-monopoles of the electric and
of the magnetic theory respectively.
This duality, known as Kim-Park duality \cite{Kim:2013cma} can be modified into a duality
involving quadratic monopoles.
There are two possibilities, depending on $k$ being even or odd.
\begin{itemize}
\item For even $k$ we add to the electric theory the monopole superpotential
\begin{equation}
\label{Weven1}
\Delta W_\mathrm{ele} = T^2_{\frac{k}{2}} + \widetilde T^2_{\frac{k}{2}-1}
\end{equation}
or equivalently
\begin{equation}
\label{Weven2}
\Delta W_\mathrm{ele} = T^2_{\frac{k}{2}-1} + \widetilde T^2_{\frac{k}{2}}
\end{equation}
\item For odd $k$
we add to the electric theory the monopole superpotential
\begin{equation}
\label{Wodd}
\Delta W_\mathrm{ele} = T^2_{\frac{k-1}{2}} + \widetilde T^2_{\frac{k-1}{2}}
\end{equation}
\end{itemize}
From now on we discuss only the case of odd $k$ and then comment on
the other case at the end.
By adding the superpotential (\ref{Wodd})
we constraint the R-charges of the monopoles and as a consequence
the one of the matter fields.
We are left with the constraint
\begin{equation}
N_f(1-\Delta) - \left(N_c-1-\frac{k-1}{2}\right) \frac{2}{k+1} = 1
\end{equation}
where $R[Q] = R[\widetilde Q] = \Delta$.
Observe that if we add the same superpotential in the dual theory
the constraint is 
\begin{equation}
N_f \left[ 1- \left( \frac{2}{k+1}-\Delta \right) \right] - \left(\widetilde N_c-1-\frac{k-1}{2}\right) \frac{2}{k+1} = 1
\end{equation}
and, at the level of the charges, this is consistent with the duality only if $\widetilde N_c = k N_f -N_c$.
This fact can be confirmed by looking at the partition function.
The identity for the Kim-Park duality is
\begin{eqnarray}
\label{KPd}
Z_{U(N_c)}(\mu;\nu;\tau;\eta)
&=& \prod_{j=0}^{k-1}  \Gamma_h \Big(\pm \frac{\eta}{2}+ \omega N_f+ (j-N_c+1) \tau - \sum_{a=1}^{N_f} \frac{\mu_a+\nu_a}{2} \Big)
\nonumber \\
&\times &
\prod_{a,b=1}^{N_f}\Gamma_h(j \tau+\mu_a +\nu_b)
Z_{U(\widetilde N_c)}(\widetilde \mu;\widetilde \nu;\tau;-\eta)
\end{eqnarray}
where we refer to appendix \ref{AppA} for the various notations.

By adding the superpotential (\ref{Wodd}) we introduce a balancing condition 
\be
\omega N_f+ \left(\frac{k-1}{2}-N_c+1\right) \tau - \sum_{a=1}^{N_f} \frac{\mu_a+\nu_a}{2} = \omega
\ee
Since
\be
\tau = \omega \Delta_A = \frac{2}{k+1} \omega
\ee
this simplifies to
\be
\label{bcadj}
 \sum_{a=1}^{N_f} (\mu_a+\nu_a) = 2 (\omega N_f - \tau N_c)
  ~.
\end{equation}
Furthermore we set $\eta = 0$.  In each of the first and the second line of \eref{KPd}, the terms in the product can be paired between $j=m$ and $j=(k-1)-m$, with $0 \leq m \leq k-1$.  In each of these two lines, there is also an unpaired term for $j=\frac{1}{2}(k-1)$. Using the identity $\Gamma_h(2\omega - x)\Gamma_h (x)=1$, it can be seen that the contributions of each pair cancel precisely, and the unpaired term gives $\Gamma_h (\omega)=1$.  Hence, we have proven that
\begin{eqnarray}
\label{quadadj}
Z_{U(N_c)}(\mu;\nu;\tau;0)
&=&
\prod_{a,b=1}^{N_f}\Gamma_h(j \tau+\mu_a +\nu_b)
Z_{U(\widetilde N_c)}(\widetilde \mu;\widetilde \nu;\tau;0)
\end{eqnarray}
with the duality map $\tilde \mu = \tau - \mu$ and 
$\tilde \nu = \tau - \nu$ and the balancing condition (\ref{bcadj}).
Observe that the even cases work in a similar manner, essentially because they leave the balancing condition (\ref{bcadj}) unchanged.

\subsubsection*{\it More general monopole superpotentials}
The above discussion can be generalised in the case in which we add to the electric theory the monopole superpotential
\begin{equation} \label{Weq}
\Delta W_{\mathrm{ele}} = T^2_{q} + \widetilde T^2_{k-1-q}
\end{equation}
and similarly to the magnetic theory the monopole superpotential
\begin{equation} \label{Wmq}
\Delta W_{\mathrm{mag}} = t^2_{q} + \widetilde t^2_{k-1-q}~.
\end{equation}

Let us first analyse the electric theory. It can be easily seen that the basic monopole operators $T_0$ and $\tilde{T}_0$ have different R-charges if $q \neq (k-1)/2$. Moreover, the $U(1)_T$ topological symmetry and the $U(1)_R$ R-symmetry is broken to a diagonal subgroup.  Let us refer to the latter as $U(1)_{R'}= U(1)_{R}-\alpha U(1)_{T}$.  Therefore,
\be
\begin{split}
R'[T_0] =R -\alpha~, \qquad
R'[\tilde{T}_0] = R+ \alpha~,
\end{split}
\ee
with
\be
R =   N_f(1-\Delta) +(N_c-1)\left( 1- \Delta_A \right) -(N_c-1), \qquad \Delta_A = \frac{2}{k+1}~.
\ee

The $R'$-charges of $T_q$ and $\tilde{T}_{k-1-q}$ can be written as follows:
\be \label{RTq}
\begin{aligned}
R'[T_q] = 1 &=  R-\alpha+ q \Delta_A \\ 
R'[\tilde{T}_{k-1-q}] = 1&= R+\alpha + (k-1-q) \Delta_A
\end{aligned} 
\ee
Solving these equations yields
\be
\alpha =  (q+1) \Delta_A-1= \frac{2q-(k-1)}{k+1}~.
\ee
This is in agreement with the above analyses for $q=(k-1)/2$ with $k$ odd, and for $q= k/2$ with $k$ even.

Similarly for the magnetic theory, we have
\be
\begin{split}
R'[t_0] = \hat{R} - \alpha~, \qquad R'[\tilde{t}_0] =\hat{R} + \alpha~,
\end{split}
\ee
with
\be
\hat{R} = N_f[1-(\Delta_A-\Delta)] +(\tilde{N}_c-1)\left( 1- \Delta_A \right) -(\tilde{N}_c-1)~,
\ee
and
\be
\begin{aligned}
R'[t_q] = 1 &=  \hat{R} - \alpha+ q \Delta_A \\ 
R'[\tilde{t}_{k-1-q}] = 1&= \hat{R}+\alpha + (k-1-q) \Delta_A~.
\end{aligned} 
\ee
Solving these equations, we obtain
\be
\tilde{N}_c = k N_f-N_c~.
\ee

We see that the sum of the equations in \eref{RTq} gives rise to the same balancing condition as \eref{bcadj}, which is independent of $q$.  
It should be emphasised that for $q\neq \frac{k-1}{2}$, the FI parameter $\xi$ in \eref{KPd} can be non-zero (in contrast with $q=\frac{k-1}{2}$).  In this case, we can pair the terms $j=m$ in the first line with $j=(k-1)-m$ in the second line, for $0 \leq m \leq k-1$.  Upon using the identity $\Gamma_h(2\omega-x)\Gamma_h(x)$=1, we see that the contribution from each pair cancels precisely.  We thus arrive at a similar relation to \eref{quadadj}:
\begin{eqnarray}
Z_{U(N_c)}(\mu;\nu;\tau; \eta)
=
\prod_{a,b=1}^{N_f}\Gamma_h(j \tau+\mu_a +\nu_b)
Z_{U(\widetilde N_c)}(\widetilde \mu;\widetilde \nu;\tau-\eta)
\end{eqnarray}
Thus, the same duality holds with the addition of \eref{Weq} and \eref{Wmq} for any $0 \leq q \leq k-1$, with a non-zero FI parameter in the partition function.


\subsection{$U(N_c)$ with a single quadratic monopole superpotential}

Here we discuss a duality between
\begin{itemize}
\item
$U(N_c)$ adjoint SQCD with 
\begin{equation}
\label{elequad1}
W = X^{k+1} + \tilde T^2_{\frac{k-1}{2}}
\end{equation}
and $k$ even, and
\item
$U(k N_f-N_c)$ adjoint SQCD with 
\begin{equation}
\label{magquad1}
W = Y^{k+1} + t^2_{\frac{k-1}{2}}
+ \sum_{j=0}^{k-1} M_j q Y^{k-1-j} \tilde q
 + \sum_{j=1}^{k-1} T_j \tilde{t}_{k-1-j}
\end{equation}
\end{itemize}
Observe that a more general duality can be constructed 
by considering a monopole superpotential $W \sim \tilde{T}^2_q$ 
with $0 \leq q \leq k-1$. Such a duality can be defined for both
even and odd $k$, and it just requires more care in the choice of the
FI (see the discussion at the end of sub-section (\ref{subsecUadj})). We will not discuss this generalization further, and leave the 
details to the interested reader.

Here we show that the duality for even $k$ summarized above can be 
obtained from the duality with quadratic monopole superpotential 
discussed in sub-section (\ref{subsecUadj}).
We consider the case with $N_f+1$ fundamentals and trigger a real mass 
flow on the partition function by considering the large $s$ limit in the relations 
\begin{equation}
\label{rmf}
\mu_{N_f+1} \rightarrow \frac{\eta}{2} + s,
\quad
\nu_{N_f+1} \rightarrow \frac{\eta}{2} - s
\end{equation}
The balancing condition (\ref{bcadj}) is modified as
\begin{equation}
\label{bcq}
\sum_{a=1}^{N_f} (\mu_a+\nu_a) + \eta
+ 2 N_c \tau = 2 \omega(N_f+1)
\end{equation}
In the dual side we further consider the Higgs flow
in the gauge sector, breaking the gauge symmetry as
$U(k (N_f+1) -N_c) \rightarrow U(k N_f -N_c) \times U(k)$.

By performing the large $s$ limit in the identity (\ref{quadadj})
we are left with the identity 
between two finite quantities, after we simplify the divergent pieces.
The subsequent analysis is very similar to that presented in \cite[sec. 4.1.1]{Amariti:2018gdc}.

On the electric side we have the partition function of $U(N_c)$ 
adjoint SQCD with superpotential (\ref{elequad1}) and effective FI 
equal to
$\Big(\frac{\eta}{2}- \omega \Big)$.
The presence of the quadratic monopole
superpotential in (\ref{elequad1}) is 
captured by the balancing condition 
(\ref{bcq}).
On the magnetic side we have two gauge sectors.
The first one corresponds to $U(k N_f-N_c)$ adjoint SQCD
and it captures the first three terms in the superpotential
 (\ref{magquad1}).
The last term in  (\ref{magquad1}) (i.e. the contribution
of the electric dressed monopoles $\tilde T_j$ acting as singlets in
the dual phase) is captured by the second integral. 
In addition there is are $j$ contributions from the 
$(N_f+1)$-th components of 
the original dressed mesons, that are massless in this dual phase, after triggering the real mass flow as in (\ref{rmf}).
The contribution of the singlets $T_j$ can be seen explicitly 
by studying the partition function associated to this 
extra gauge sector and these $j$ singlets arising from the original meson.
We have
\begin{eqnarray}
\label{monm}
&&
\prod_{j=0}^{k-1} 
\Gamma_h(\eta+ j \tau )
\int \prod_{c=1}^{k} \mathd \sigma_c
\,
\mathe^{\mathi \pi (\eta-2 \omega)}
\,
\Gamma_h\left(\tau - \frac{\eta}{2}\pm\sigma_c\right)
\prod_{c<d} \frac{\Gamma_h(\pm(\sigma_c - \sigma_d) + \tau)}
{\Gamma_h(\pm(\sigma_c - \sigma_d))}
\nonumber \\
=
&&
\prod_{j=0}^{k-1}
\Gamma_h(\eta+ j \tau )
\Gamma_h(\eta-(j+1) \tau)
\Gamma_h(2 \omega-(j+1) \tau)
\Gamma_h((2-j)\tau-\eta)
\nonumber \\
=
&&
\prod_{j=0}^{k-1}
\Gamma_h(\eta-(j+1) \tau)
=
\prod_{j=0}^{k-1}
\Gamma_h(\eta-(k-j)\tau)
\end{eqnarray}
where we have evaluated this integral by using \cite[{Theorem 5.6.8}]{VdB}.
We can show that (\ref{monm}) 
 corresponds to the electric monopole 
 by applying the balancing condition (\ref{bcq}):
\begin{equation}
\label{mch}
\eta - (k-j) \tau 
=
\left(\frac{\eta}{2} - \omega \right)
+[j-k-N_c] \tau+\omega (N_f+2) 
- \frac{1}{2} \sum_{a=1}^{N_f} (\mu_a + \nu_a)
\end{equation}
In order to see that this combination
captures the global charges of the 
dressed monopoles $T_j$ acting as singlets in the 
last sum of the superpotential (\ref{magquad1}) we have to 
shift the effective FI as 
$\eta \rightarrow \eta+2 \omega$.
In this way the FI is chosen canonically and we can 
simply read the global charges from the combination 
of the real masses in  (\ref{mch}).
After the shift and some rearranging the RHS of (\ref{mch}) becomes
\begin{equation}
\label{mch2}
\frac{\eta}{2} +\omega N_f -(N_c-j-1) \tau 
- \frac{1}{2} \sum_{a=1}^{N_f} (\mu_a + \nu_a)
\end{equation}
From this relation we can see that this fields has topological charge
$+1$, axial mass $-N_f$ and R-charge
\begin{equation}
\Delta_j = N_f(1-\Delta_Q) -\Delta_X (N_c-j-1)
\end{equation}
where $\Delta_Q$ are the charges of the electric fundamentals $Q$ 
and antifundamentals $\widetilde Q$ (with $\Delta_Q = 
\Delta_{\widetilde Q}$)
 and $\Delta_X$ is the R-charge of the electric adjoint field $X$. 
 This shows that the expression in (\ref{mch2}) 
is the combination of masses and charges
expected for the (dressed) electric monopoles.

Observe that from this duality we can further flow to 
the identity (\ref{KPd}) by triggering a further real mass
flow. This provides a further consistency check
of the duality. We leave the details of this calculation
to the interested reader.

\subsection{$U(N_c)_\kappa$ with  quadratic monopole superpotentials}

It is also possible to study an RG flow leading to a
duality involving CS matter theories.
This is done by turning on the real masses
for the fundamentals and shifting the scalars
$\sigma_i$ and the FI as
\begin{eqnarray}
\begin{array}{ll}
\mu_a \rightarrow \mu_a  - \kappa s  & \quad a=1,\dots,N_f-\kappa \\
\mu_a \rightarrow \mu_a + (2 N_f-\kappa) s 
&\quad a=N_f-\kappa+1,\dots,N_f \\
\nu_a \rightarrow \nu_a  + \kappa s  &\quad a=1,\dots,N_f \\
\eta \rightarrow \eta - 2 N_f \kappa s \\
\sigma_i \rightarrow \sigma_i + \kappa s& \quad  i=1,\dots,N_c \\
\widetilde \sigma_i \rightarrow \sigma_i - \kappa s& \quad  i=1,\dots,
\widetilde N_c \\
\end{array}
\end{eqnarray}
where $\widetilde \sigma_i $ is the shift of the scalar in the dual vector multiplet. The real masses in the dual theory can be read from
the duality map as usual.
We can study the real mass flow by computing the large $s$ limit 
on the partition function. 
We find the following identity 
\begin{eqnarray}
\label{CTphi}
Z_{U(N_c)_{\frac{\kappa}{2}}}(\mu;\nu;\tau;\eta_{\mathrm{ele}})
&=&
\mathe^{\frac{\mathi \pi}{2} \phi}
\prod_{a=1}^{N_f-\kappa}
\prod_{b=1}^{N_f}
\prod_{j=0}^{k-1}
\Gamma_h(\mu_a + \nu_b + j \tau)
\nonumber \\
&\times&
Z_{U(k N_f-N_c)_{-\frac{\kappa}{2}}}(\tau-\mu;\tau-\nu;\tau;\eta_{\mathrm{mag}})
\end{eqnarray}
where the electric and the magnetic FI  in (\ref{CTphi})  are
\begin{eqnarray}
\eta_{\mathrm{ele}} &=& - 2 \big(\sum _{a=1}^{N_f-\kappa} \mu_a
-\sum _{b=1}^{N_f } \nu_b + \eta -\omega ( \kappa  +2) \big),
\nonumber \\
\eta_{\mathrm{mag}} &=& - 
2 \big(
\sum _{a=1}^{N_f-\kappa} \mu_a
-
\sum _{b=1}^{N_f } \nu_b+ \eta - \kappa  \tau + \omega ( \kappa-2)\big)
\end{eqnarray}
and the phase $\phi$ in (\ref{CTphi})  is
\begin{eqnarray}
\phi 
&=&
k\bigg(
\kappa  \sum _{b=1}^{N_f} \nu _b^2
-2 \big(\sum _{b=1}^{N_f} \nu _b-\sum _{a=1}^{N_f-\kappa } \mu _a\big) \big(\sum _{b=1}^{N_f} \nu _b+\tau  N_c-\omega  N_f\big)-2 \kappa  \tau  \sum _{b=1}^{N_f} \nu _b
\bigg)
\nonumber \\
&-&
\frac{1}{3} k \left(\kappa  \tau  \left(3 \tau  N_c+(k-4) \omega  N_f\right)-\tau  \omega +13 \omega ^2\right)-k\eta ^2 +4 \eta  k \omega
\end{eqnarray}
This is compatible with a duality between
\begin{itemize}
\item
$U(N_c)_\frac{\kappa}{2}$ adjoint SQCD with 
$N_f-\kappa$ fundamentals and $N_f$ 
anti-fundamentals and  superpotential 
\begin{equation}
\label{elequad1}
W = X^{k+1} + \tilde T^2_{\frac{k-1}{2}}
\end{equation}
and $k$ even, and
\item
$U(k N_f-N_c)_{-\frac{\kappa}{2}}$ adjoint SQCD 
with 
$N_f-\kappa$ fundamentals and $N_f$ 
anti-fundamentals and  superpotential 
\begin{equation}
\label{magquad1}
W = Y^{k+1} + t^2_{\frac{k-1}{2}}
+ \sum_{j=0}^{k-1} M_j q Y^{k-1-j} \tilde q
\end{equation}
\end{itemize}
This duality generalises that of \cite[sec. 8.1]{Benini:2017dud} for the linear monopole superpotential and that of \cite[sec. 3.2.3]{Amariti:2018gdc} for the quadratic monopole superpotential.

\subsection{$U(N_c)$ with  linear and quadratic monopole superpotentials}

Here we discuss a duality between
\begin{itemize}
\item
$U(N_c)$ adjoint SQCD with 
\begin{equation}
\label{elequad2}
W = X^{k+1} + \tilde T^2_{\frac{k-1}{2}} + T_0
\end{equation}
and $k$ even, and
\item
$U(k (N_f-1)-N_c)$ adjoint SQCD with 
\begin{equation}
\label{magquad2}
W = Y^{k+1}  + \sum_{j=0}^{k-1} M_j q Y^{k-1-j} \tilde q
 +   \tilde t^2_{\frac{k-1}{2}}+ t_0
\end{equation}
\end{itemize}
Again a more general duality can be constructed, including also the $k$ odd case, by considering a monopole superpotential $W \sim \tilde{T}_q^2$ 
(or $W \sim T_q^2$) with $0 \leq q \leq k-1$. We will not further
discuss this generalization here.

Here we provide an evidence of this duality, showing that
it can be obtained from a duality discussed in \cite{Amariti:2018wht} involving
$U(N_c)$ adjoint SQCD with 
$W = X^{k+1} + T_0$
and
$U(k (N_f-1)-N_c)$ adjoint SQCD with 
$W = Y^{k+1}  + \sum_{j=1}^{k-1} M_j q Y^{k-1-j} \tilde q
 + t_0$.
 
 Our argument will be based on the matching of the 
 partition functions. We start from the relation derived in 
 \cite{Amariti:2018wht} 
\begin{equation}
\begin{split}
\label{idZ}
Z_{U(N_c)} (\mu;\nu;\tau;\omega -\eta) 
&= \prod_{j=0}^{k-1} 
\Gamma_h(2 \eta+ \tau j)
\prod_{a,b=1}^{N_f} \Gamma_h(\mu_a + \nu_b + j \tau) \times\\
& \qquad Z_{U(k (N_f-1)-N_c)} (\tau-\mu;\tau - \nu;\tau;\tau-\omega -\eta) 
\end{split}
\end{equation}
Note that in the $U(N_c)$ theory the FI parameter is taken to be $\omega- \eta$.
This identity is valid provided the  condition 
\begin{equation}
\label{bc222}
\sum_{a=1}^{N_f} (\mu_a +\nu_a) + \eta = \omega(N_f-1)- \tau(N_c-1)
\end{equation}
on the parameters is imposed.
Next we add the quadratic superpotential $W \sim \tilde{T}^2_\frac{k-1}{2}$ on the electric side of the duality.
It corresponds to fix $\eta = \frac{\tau}{2}$, due to the fact that
\be
1= R[\tilde{T}_\frac{k-1}{2}]=R[\tilde{T}_0] + \left( \frac{k-1}{2} \right) \Delta_A = R[\tilde{T}_0] +(1-\Delta_A) ~.
\ee
On the magnetic side the effect of this deformation can be argued 
by looking at the partition function (\ref{idZ}). The net effect consists
of giving a holomorphic  mass to the singlets associated to the monopoles of the electric theory:
\begin{equation}
\prod_{j=0}^{k-1} 
\Gamma_h(2 \eta+ \tau j)
=
\prod_{j=0}^{k-1} 
\Gamma_h(\tau (j+1))
=
\Gamma_h(\tau)
\dots
\Gamma_h(k \tau )
=
1
\end{equation}
We are then left with the identity 
\begin{equation}
\label{idZ2}
\begin{split}
Z_{U(N_c)} \left(\mu;\nu;\tau;\omega -\frac{\tau}{2} \right) 
& = \prod_{j=0}^{k-1} 
\prod_{a,b=1}^{N_f} \Gamma_h(\mu_a + \nu_b + j \tau) \times \\
&\qquad Z_{U(k (N_f-1)-N_c)} \left(\tau-\mu;\tau - \nu;\tau;\frac{\tau}{2}-\omega\right) 
\end{split}
\end{equation}
with the balancing condition 
\begin{equation}
\label{bc223}
\sum_{a=1}^{N_f} (\mu_a +\nu_a)  = \omega(N_f-1)- \tau\left(N_c-\frac{1}{2}\right)
\end{equation}
The relation (\ref{idZ2}) together with the balancing condition (\ref{bc223}) represents the matching of the partition function for the duality summarized at the beginning of this subsection.
Observe that the presence of the quadratic monopole in the magnetic superpotential, $W \sim \tilde{t}^2_{\frac{k-1}{2}}$ can be argued because it is consistent with the constraints on the global charges given by (\ref{bc223}).

\subsection{$USp(2N_c)$ gauge group}
The above duality can easily be generalised to theories with symplectic gauge groups.  We propose the duality between the following theories:
\paragraph{Theory A.} The $USp(2N_c)$ gauge theory with $2N_f$ fundamentals $Q^a$, an antisymmetric traceless chiral multiplet $A$, and a superpotential
\be
W = \tr A^{k+1} + T_q^2~,
\ee
where $T_q$ is the dressed monopole operator
\be
T_q =Y \tr(A^q)
\ee 
with $q$ integer and $Y$ the basic monopole operator of theory A. 
\paragraph{Theory B.}  The $USp(2\tilde{N}_c)$ gauge theory with $2N_f$ fundamentals, an antisymmetric traceless chiral multiplet $a$, singlets $M_j =Q^a Y^j Q^b$ ($j=0,\ldots, 2k$), singlets $T_j =T_0 \tr X^j$, and a superpotential
\be
W = \tr a^{k+1} + \sum_{j=0}^{k-1} M_{k-j-1}q a^j q + t_q^2~. 
\ee
where $t_q$ is the dressed monopole operator 
\be
t_q =\tilde{Y} \tr(a^q)~.
\ee 
with $\tilde{Y}$ the basic monopole operator of theory B.

We will see that the duality holds provided that
\be \label{condUSp}
\tilde{N}_c = (N_f-1)k - N_c~, \qquad q = \frac{1}{2}(k-1)~.
\ee
In order for $q$ to be an integer, $k$ has to be odd.  However, if $k$ is even, $q$ is half-odd-integral and we need to redefine the dress monopole operator $T_q$ and $t_q$.  One possibility is to define\footnote{The determinant of $A$ is related to the trace of a power of $A$ by Newton's identities.  Note that since $A$ is an anti-symmetric matrix, the trace of an odd power of $A$ is zero.  Thus, for example, if $A$ is a two by two matrix, we have $\det A = -\frac{1}{2} \tr(A^2)$; and if $A$ is a four by four matrix, we have $\det A = \frac{1}{8} (\tr(A^2))^2-\frac{1}{4} \tr(A^4)$.}
\be \label{redefTq}
T_q = Y (\det A)^{\frac{q}{2N_c}}~, \quad t_q = \tilde{Y} (\det a)^{\frac{q}{2\tilde{N}_c}}~.
\ee
for $q$ either integral or half-odd-integral.

From the superpotentials, we see that the R-charges of $A$ and $a$ are equal to 
\be
\Delta_A  = \frac{2}{k+1}~.
\ee
In order to see the first equality of \eref{condUSp}, we consider the R-charges of the monopole operator $T_q$ and $t_q$:
\be
\begin{split}
1=R[T_q] &= 2 N_f (1 - r) + q \Delta_A + (1 - \Delta_A) 2 (N_c - 1) - 2 N_c  \\
1=R[t_q] &= 2 N_f (1 - (\Delta_A-r)) + q \Delta_A + (1 - \Delta_A) 2 (\tilde{N}_c - 1) - 2 \tilde{N}_c ~.
\end{split}
\ee
Solving these equation, we obtain
\be
\tilde{N}_c = (N_f-1)k - N_c + \left[ q -\frac{1}{2}(k-1) \right]~.
\ee
The monopole superpotential in theory $A$ gives rise to the balancing condition:
\be \label{BCUSp}
\begin{split}
2\omega N_f  +   \left[q - 2(N_c-1) \right] \tau -\sum_{a=1}^{2N_f} \mu_a =3 \omega~.
\end{split}
\ee

The identity for the duality without monopole superpotentials is given by \cite[(5.5)]{Amariti:2015vwa}:
\be
\begin{split}
Z_{USp(2N_c)} (\mu; \tau) &= \prod_{j=0}^{k-1} \,\, \prod_{a< b} \Gamma_h( \mu_a +\mu_b + j \tau) Z_{USp(2\tilde{N}_c)} (\tau-\mu; \tau)  \times \\
& \quad \prod_{j=0}^{k-1} \Gamma_h\left( -2\omega + 2 \omega N_f + [j-2(N_c-1)]\tau -\sum_{a=1}^{2N_f} \mu_a \right)~.
\end{split}
\ee

Let us assume that $k$ is odd.  We see that the terms in the product in the second line of the above equation can be paired between $j=m$ and $j=(k-1)-m$, with $0 \leq m \leq k-1$.  There is an unpaired term for $j=\frac{1}{2}(k-1)$.  The argument of $\Gamma_h$ for each pair adds up to
\be
-4\omega + 4 \omega N_f + 2 \left[ \frac{1}{2}(k-1)-2(N_c-1) \right]\tau -2\sum_{a=1}^{2N_f} \mu_a~.
\ee
Upon using the balancing condition \eref{BCUSp}, with $q= \frac{1}{2}(k-1)$, the above expression becomes $2\omega$.  We can then use the identity $\Gamma_h(2 \omega-x) \Gamma(x)=1$ to cancel the contribution of each pair.  On the other hand, the argument of $\Gamma_h$ for the unpaired term is
\be
-2\omega + 2 \omega N_f + \left[ \frac{1}{2}(k-1) -2(N_c-1) \right]\tau -\sum_{a=1}^{2N_f} \mu_a  = \omega~,
\ee
we we have used again the balancing condition \eref{BCUSp}, with $q= \frac{1}{2}(k-1)$.  Since $\Gamma_h(\omega) = 1$, we obtain
\be
Z_{USp(2N_c)} (\mu; \tau) = \prod_{j=0}^{k-1} \,\, \prod_{a< b} \Gamma_h( \mu_a +\mu_b + j \tau) Z_{USp(2\tilde{N}_c)} (\tau-\mu; \tau) ~, \quad q=  \frac{1}{2}(k-1)~.
\ee
We thus establish the duality between theories $A$ and $B$, with the parameters $q=  \frac{1}{2}(k-1)$ and thus $\tilde{N}_c = (N_f-1)k -N_c$, along with the duality map $\tilde{\mu} = \tau-\mu$, $\tilde{\nu} =\tau-\nu$.

In the case in which $k$ is even, we see that there is no unpaired term and the contributions from each pair cancel precisely.   However, in this case, $q$ takes a half-odd-integral value, and so the dressed monopole operators have to be redefined as, for example, in \eref{redefTq}.

\section{Conclusions}

In this paper we discussed 3d $\mathcal{N}=2$ dualities in the
presence 
of quadratic monopole superpotentials.
In the first part of the paper we provided a brane picture of such
dualities for SQCD with symplectic and unitary gauge groups.
The basic observation is that these dualities can be obtained
by T-duality on the 4d picture in the presence of orientifolds, as discussed in \cite{Amariti:2015yea,Amariti:2015mva,Amariti:2016kat,Amariti:2017gsm}. The new ingredient that allowed us here to generalize the construction to the cases with quadratic monopole superpotentials 
corresponds to consider also twisted affine compactifications.
The twist is due to an outer automorphism of the gauge algebra and it 
implies that after T-duality we can have all the possible pairs involving
$\Ot^-$, $\Ot^+$ and $\Ott^-$, 
acting on the compact direction.
This provides a classification scheme for the  3d $\mathcal{N}=2$.
In this way we obtain also a new duality for the 
unitary case, with a linear  and a quadratic monopole superpotential.
This duality has been checked against the partition function as well.
In the second part of the paper we used similar arguments on the partition function to construct new dual pairs for dualities with tensorial matter, adjoint for the unitary case and antisymmetric for the symplectic one.

In the analysis we left some open question on which we would like to come back in the future.
First, we did not discuss the orthogonal case. This corresponds to 
consider the gauge theory living on $\Ot^{-}$ on $\Ott^{-}$
planes.
The monopole superpotential in these cases has a more intricate structure, because it involves the monopole $Y_{Spin}$ or the
monopole $Y \propto \mathe^{\Sigma_1}$. The two are related as 
$Y^2 = Y_{Spin}$ and while $Y_{Spin}$ exists for both $SO(N_c)$ and $Spin(N_c)$ gauge groups, the monopole $Y$ can be defined only 
for $SO(N_c)$. 
It should then be necessary to further study these models from the perspective of their global properties.

In the analysis of the unitary theories we also used a \emph{caveat},
summarized in the presence of two different boundary conditions at the positions of the orientifolds.
More concretely,  the boundary condition corresponding to the $\Ot^-$ at $x_3 =\pi R$ gives rise to a term in the superpotential that is not sensitive to the radius to the circle and can be sent to zero upon shrinking the radius of the circle.  However, the other boundary condition $\Ot^+$ at $x_3 = 0$ gives rise to a term in the superpotential that is sensitive to the radius of the circle. It should be interesting further elaborate on 
this difference. 

Another interesting analysis that we did not perform consists in 
reproducing the dualities studied in section \ref{plaw} from the 
brane picture. In such cases there will be singular configurations, based
on the presence of stacks of NS branes.
These configurations were studied in \cite{Amariti:2015mva} for the reduction of 
4d dualities to 3d.  It should be possible to reproduce on such brane configurations the quadratic monopole superpotentials and the dualities discussed here in section \ref{plaw}.

Our analysis may be also generalized by considering $U(N_c)$ SQCD with two  adjoints.
The duality was derived in \cite{Hwang:2018uyj}, inspired by the 4d duality of \cite{Brodie:1996vx}.
It should be possible to see if a quadratic monopole deformation can be added to this duality and if it gives raise to a new IR duality.
Other generalizations to the dualities with tensorial matter studied in 
\cite{Amariti:2015mva} are expected as well.

We would like to conclude by observing a problem that appears in the unitary case with $N_f=N_c=1$.
In this case the quadratic monopole deformation gives raise to a divergent partition function. This seems to be the case also when deforming the SQCDA/XYZ duality discussed in \cite{Nieri:2018pev}, by adding the deformations
(\ref{Weven1}), (\ref{Weven2}) or (\ref{Wodd}) .
This corresponds to a flat direction in the Coulomb branch and the 
partition function argument cannot be used in such cases.
Further checks are necessary to check the duality in this case.

\acknowledgments
We thank Domenico Orlando, Sergio Benvenuti and Susanne Reffert for valuable comments.  The work of LC is supported in part by Vetenskapsr\r{a}det under grant \#2014-5517, by the STINT grant, and by the grant  ``Geometry and Physics"  from the Knut and Alice Wallenberg foundation.

\appendix
\section{Three sphere partition functions}
\label{AppA}

In this appendix we provide some useful formulae
for the 3d $\mathcal{N}=2$ partition function on a 
squashed three sphere, used in the body of the paper.
We refer to 
\cite{Kapustin:2009kz,Jafferis:2010un,Hama:2010av,Hama:2011ea} 
for the original derivation in localization.

The partition function of a gauge theory is an integral
over the Cartan of the gauge group.
This is parameterized by the real scalars $\sigma$, representing the dynamical real scalar in the $\mathcal{N}=2$ vector multiplet.
There are classical contributions, corresponding to the FI and to the CS terms in the action, and quantum contributions, represented by the one-loop 
determinants of the gauge and matter fields.
These one-loop determinants can be formulated in terms of 
hyperbolic Gamma functions, $\Gamma_h(\sigma)$ (see \cite{VdB} for a 
definition and \cite{Benini:2011mf} for a physical interpretation) 
\begin{equation}
  \label{GG}
  \Gamma_h(z;\omega_1,\omega_2) \equiv
  \Gamma_h(z)\equiv 
  \mathe^{
    \frac{\mathi \pi}{2 \omega_1 \omega_2}
    ((z-\omega)^2 - \frac{\omega_1^2+\omega_2^2}{12})}
  \prod_{\alpha=0}^{\infty} 
  \frac
  {1-\mathe^{\frac{2 \pi \mathi}{\omega_1}(\omega_2-z)} \mathe^{\frac{2 \pi \mathi \omega_2 \alpha}{\omega_1}}}
  {1-\mathe^{-\frac{2 \pi \mathi}{\omega_2} z} \mathe^{-\frac{2 \pi \mathi \omega_1 \alpha}{\omega_2}}}.
\end{equation}
where the argument $z$ can be further refined by adding the contributions of the scalars in the background vector multiplets.
The purely imaginary parameters $\omega_1 \equiv \mathi b$ and $\omega_2 \equiv \mathi b^{-1}$ are defined in terms of the real squashing parameter $b$ of the ellipsoid $S_b^3$, and  
$2 \omega \equiv \omega_1+\omega_2$.

There are two types of background symmetries, flavor and R-symmetries.
We can turn on a collective  background scalar $\mu$ for the first and 
$\Delta$ for the second, where $\Delta$ is the R-charge, equivalent to the mass dimension in three dimensions.
We then define a holomorphic combination $\mu + \omega \Delta$.  In other words, we can count the contribution of the 
R-symmetry by turning on an imaginary part for the real masses.
We now restrict to the partition functions of interest in the paper, that
are $U(N_c)$ SQCD with an adjoint and $USp(2N_c)$ SQCD
with a traceless antisymmetric. 

In the first case the partition function can be written as
\begin{eqnarray}
\label{formgen}
Z_{U(N_c)_\kappa}(\mu;\nu;\tau;\eta)
&
=
&
\frac{1}{|W|} 
\int \prod_{i=1}^{N_c} \mathd \sigma_i \;
\mathe^{\mathi \pi \kappa \sigma_i^2 +\mathi \pi \eta \sigma_i}
\prod_{a=1}^{N_f}
\Gamma_h(\mu_a +\sigma_i,\nu_a-\sigma_i)
\nonumber \\
&\times&
\prod_{1\leq i<j \leq N_c}
\frac{\Gamma_h(\pm (\sigma_i - \sigma_j)+\tau)}
{\Gamma_h(\pm (\sigma_i - \sigma_j))}
\end{eqnarray}
where we used the shorthand notations
$\Gamma_h(x\pm y) \equiv \Gamma_h(x +  y) \Gamma_h(x -  y) $
and $\Gamma_h(x,y) = \Gamma_h(x)\Gamma_h(y)$.
The arguments in the LHS of (\ref{formgen}) refer respectively
to the real masses of the fundamentals ($\mu$), of the anti-fundamentals ($\nu$), of the adjoint ($\tau$) and the FI ($\eta$). 
$|W|$ is the order of the Weyl group of $U(N_c)$.
We also introduced a CS level $\kappa$ in (\ref{formgen}). When considering 
cases with vanishing CS we omit the $\kappa$-dependence.

For $USp(2N_c)$ with an antisymmetric we have
\begin{eqnarray}
\label{formgen2}
Z_{USp(2N_c)}(\mu;\tau)
&
=
&
\frac{1}{|W|} 
\int \prod_{i=1}^{N_c} \mathd \sigma_i 
\prod_{a=1}^{2N_f}
\Gamma_h( \pm \mu_a +\sigma_i)
\nonumber \\
&\times&
\prod_{1\leq i<j \leq N_c}
\frac{\Gamma_h(\pm \sigma_i \pm \sigma_j+\tau)}
{\Gamma_h(\pm  \sigma_i \pm \sigma_j))}
\prod_{i=1}^{N_c} 
\frac{1}{\Gamma_h(\pm 2 \sigma_i)}
\, ,
\end{eqnarray}
where the arguments in the LHS of (\ref{formgen2}) refer respectively
to the real masses of the fundamentals ($\mu$) and of the antisymmetric ($\tau$). 
In this case we omitted possible CS terms in 
(\ref{formgen2})  because they do not play any relevant role in our discussion.

The real mass flows discussed on the field theory side 
correspond to the limit 
\begin{equation}
\label{intout}
\lim_{x\rightarrow \infty} \Gamma_h(x) = \mathe^{\frac{\mathi \pi}{2}(x-\omega)^2}
\, 
.
\end{equation}
A real mass flow interpolating two dualities 
can be studied on the partition function by 
 computing the limit (\ref{intout}) on both sides of an identity 
between the partition functions of the dual phases.
In order to reproduce the IR duality one has to pay attention
in canceling the divergent contributions among the two sides 
of the identity.

\bibliographystyle{ytphys}
\bibliography{ref}

\end{document}